\documentclass[twocolumn,english,aps,preprintnumbers]{revtex4}
\pdfoutput=1
\usepackage[T1]{fontenc}
\usepackage[latin9]{inputenc}
\usepackage{prettyref}
\usepackage{textcomp}
\usepackage{amsmath}
\usepackage{graphicx}
\usepackage{amssymb}


\usepackage{babel}

\begin{document}

\title{Bose-Einstein condensation of paraxial light%
%\thanks{
%Financial support from the Deutsche Forschungsgemeinschaft within
%the focused research unit FOR557 and under contract WE1748/09 is
%acknowledged.
%Grants or other notes \%about the article that should go on the
%front page should be \%placed here. General acknowledgments should
%be placed
%at the end of the article.%
%}
}

\author{J. Klaers, J. Schmitt,  T. Damm,  F. Vewinger and
M. Weitz}
\affiliation{Institut f\"ur Angewandte Physik, Universit\"at Bonn, Wegelerstr. 8, 53115 Bonn, Germany.\\
Tel.: +49-228-734836\\
Fax: +49-228-734835\\
\email{klaers@iap.uni-bonn.de}}
%\emph{Present address:} of F. Author \% if needed \and S. Author
%\at second address}

%\date{Version: \today, final proof}
%\date{Received: date / Accepted: date}
\date{\today}

\begin{abstract}
Photons, due to the virtually vanishing pho\-ton-photon
interaction, constitute to very good approximation an ideal Bose
gas, but owing to the vanishing chemical potential a (free) photon
gas does not show Bose-Einstein condensation. However, this is not
necessarily true for a lower-dimensional photon gas. By means of a
fluorescence induced thermalization process in an optical
microcavity one can achieve a thermal photon gas with freely
adjustable chemical potential. Experimentally, we have observed
thermalization and subsequently Bose-Einstein condensation of the
photon gas at room temperature. In this paper, we give a detailed
description of the experiment, which is based on a dye-filled
optical microcavity, acting as a white-wall box for photons.
Thermalization is achieved in a photon number-conserving way by
photon scattering off the dye molecules, and the cavity mirrors
both provide an effective photon mass and a confining potential -
key prerequisites for the Bose-Einstein condensation of photons.
The experimental results are in good agreement with both a
statistical and a simple rate equation model, describing the
properties of the thermalized photon gas.
\end{abstract}

 \maketitle

\setcounter{tocdepth}{2}
 \tableofcontents

\section{Introduction\label{sec:Introduction}}

Photons have played a vital role for the experimental realization
of Bose-Einstein condensation (BEC) \cite{Einstein:1925p3} in
ultracold atomic gases
\cite{Anderson:1995,Davis:1995p2059,Bradley:1997,Jochim:2003p2017,Greiner:2003p1410,Leggett:2001}.
To cool down matter to sufficiently low temperatures,
fundamentally new cooling and trapping techniques based on the
interaction with light had to be developed. These experimental
techniques were rendered possible only by the invention of the
laser and even between the birth of the laser and the experimental
realization of Bose-Einstein condensation in atomic gases more
than thirty years elapsed.

Considering their bosonic nature it seems natural to ask if
photons can undergo a BEC in their own right. It is known that a
similar phenomenon occurs, when a mode in a laser resonator is
macroscopically occupied above the lasing threshold. The
oscillation build-up of a laser mode is however not a
thermodynamic phase transition in a conventional sense. Neither
the state of the light field is in thermal equilibrium nor the
lasing threshold can be determined by thermodynamic
considerations. To some extent a laser is even a prime example for
a system in non-equilibrium: Typically laser gain in a medium is
only achieved if the transition at the lasing wavelength is
inverted, which can only be achieved under non-equilibrium
conditions using external pumping \cite{Siegman,Milonni,
ScullyZubairy}. The question, whether a BEC of photons in a strict
sense is possible, typically is negated with reference to black
body radiation \cite{Planck:1901p2106,Bose:2005p2098}
\nobreakdash- presumably the most omnipresent Bose gas at all
\cite{Huang:StatisticalMechanics1987,Ketterle:EnricoFermi1999}.
For black body radiation not only the spectral distribution of the
photon energies depends on temperature, but also their total
number. Temperature and photon number cannot be tuned
independently. If the temperature of the photon gas is lowered
then also the photon number will decrease according to the
Stefan-Boltzmann law, preventing the onset of Bose-Einstein
condensation.

The last years have seen increasing effort in investigating
equilibrium processes that lead to macroscopically occupied
photonic states. It is clear that such processes have to conserve
the photon number. Presumably the first process to be considered
was Compton scattering of x-ray photons in plasmas
\cite{Zeldovich:1969p1287}. Later theoretical work considered the
use of cavities for photon condensation, but did not give a route
for reaching the thermalization \cite{Mueller:1986}. In an
interesting work, photon-photon scattering in a nonlinear medium
was investigated
\cite{Chiao:PhysicalReviewA1999,Chiao:OpticsCommunications2000,Mitchell:2000p1910,Bolda:2001p514,McCormick:OpticsExpress2002,McCormick:Thesis2003,Navez:2003p673,Chiao:arXiv:physics/0309065,Martinez:2006p476,Seaman:2008p713}.
Here photons are expected to thermalize due to four-wave mixing
which can be interpreted as collisions between two photons
\nobreakdash- quite similar to binary collisions of atoms.
However, this ansatz could not be verified experimentally. The
photon-photon interactions in the experimental systems
investigated up to now seem to be too weak to achieve a
thermalization \cite{Mitchell:2000p1910}. In other work,
condensation of classical waves in nonlinear or disordered media
has been investigated, though the systems are non-thermodynamic
\cite{Connaughton:2005,Conti:2008,Aschieri:2011,Weill:2010p16520}.
Experiments on exciton polariton condensation
\cite{Kasprzak:2006p444,Balili:2007p1342,Kasprzak:2008p1293,Kasprzak:2008p716,Lagoudakis:NaturePhysics2008,Amo:NaturePhysics2009}
can also be put into the broader context of light condensation.
These systems are characterized by a strong coupling of photonic
and exitonic degrees of freedom which leads to new energy
eigenstates, the exciton polaritons. Due to their excitonic part,
which is coherently coupled to the photonic part, the polaritons
perform binary collisions, which can lead to a thermalization of
the polariton gas.

Recently we have observed a photon number conserving
thermalization process of a two-dimensional photon gas in a dye
microresonator \cite{Klaers:2010p2137}. Here a thermal contact
between photons and a dye solution working as a heat bath is
achieved. This leads to a thermalization of the transversal photon
state. Signatures of this thermalization process are Bose-Einstein
distributed photon energies and a spatial relaxation of the photon
gas induced by an effective trapping potential. Once the  phase
space density approaches a value of order unity Bose-Einstein
condensation to the transversal ground mode can be observed
\cite{Klaers:2010}. It is one of the purposes of this paper to
give a more detailed account on the experiments we have performed
so far. In section \ref{sec:ExperimentalScheme} we describe the
basic ideas behind the experiment, together with a rate equation
model describing the thermalization. In section
\ref{sec:ExperimentalSetup} the experimental apparatus is
described. The experimental results obtained concerning the
thermalization of the photon gas are presented in section
\ref{sec:thermalization}, followed by a detailed account on the
experiments on Bose-Einstein condensation of photons in section
\ref{sec:BEC}. We conclude the paper with an outlook on possible
future directions for the photonic BEC.

\section{Experimental scheme\label{sec:ExperimentalScheme}}

\begin{figure}
a)

\includegraphics[width=0.35\textwidth]{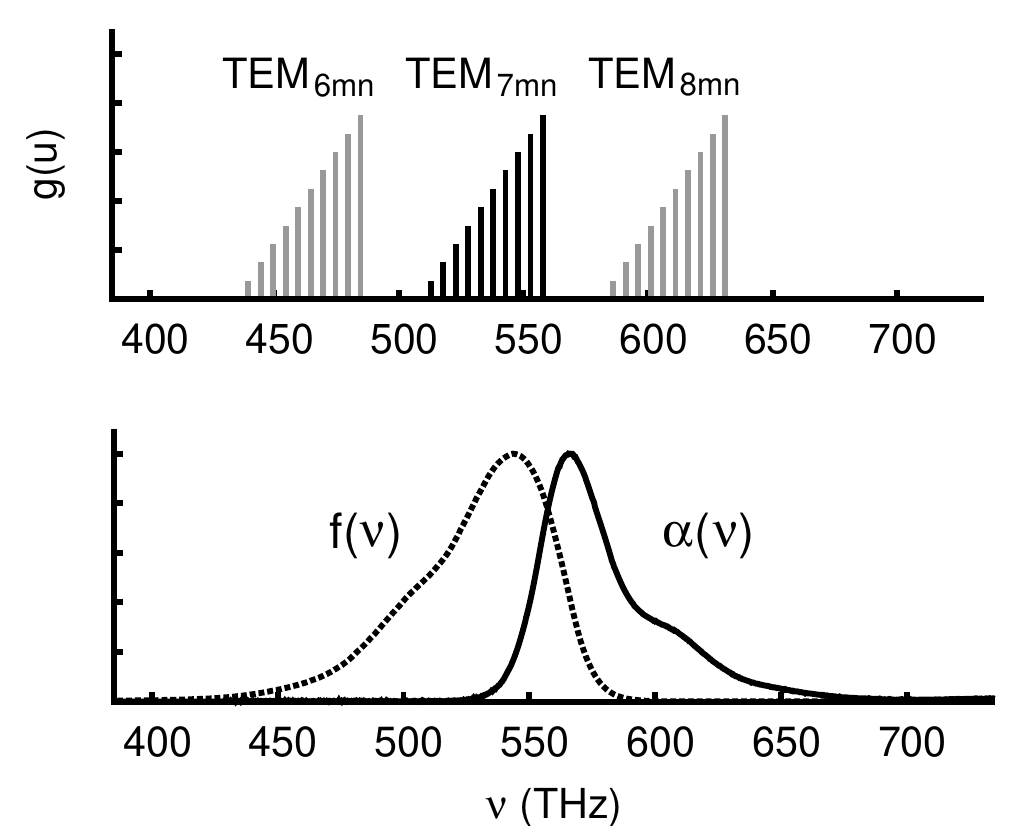}

b)

\includegraphics[width=0.40\textwidth]{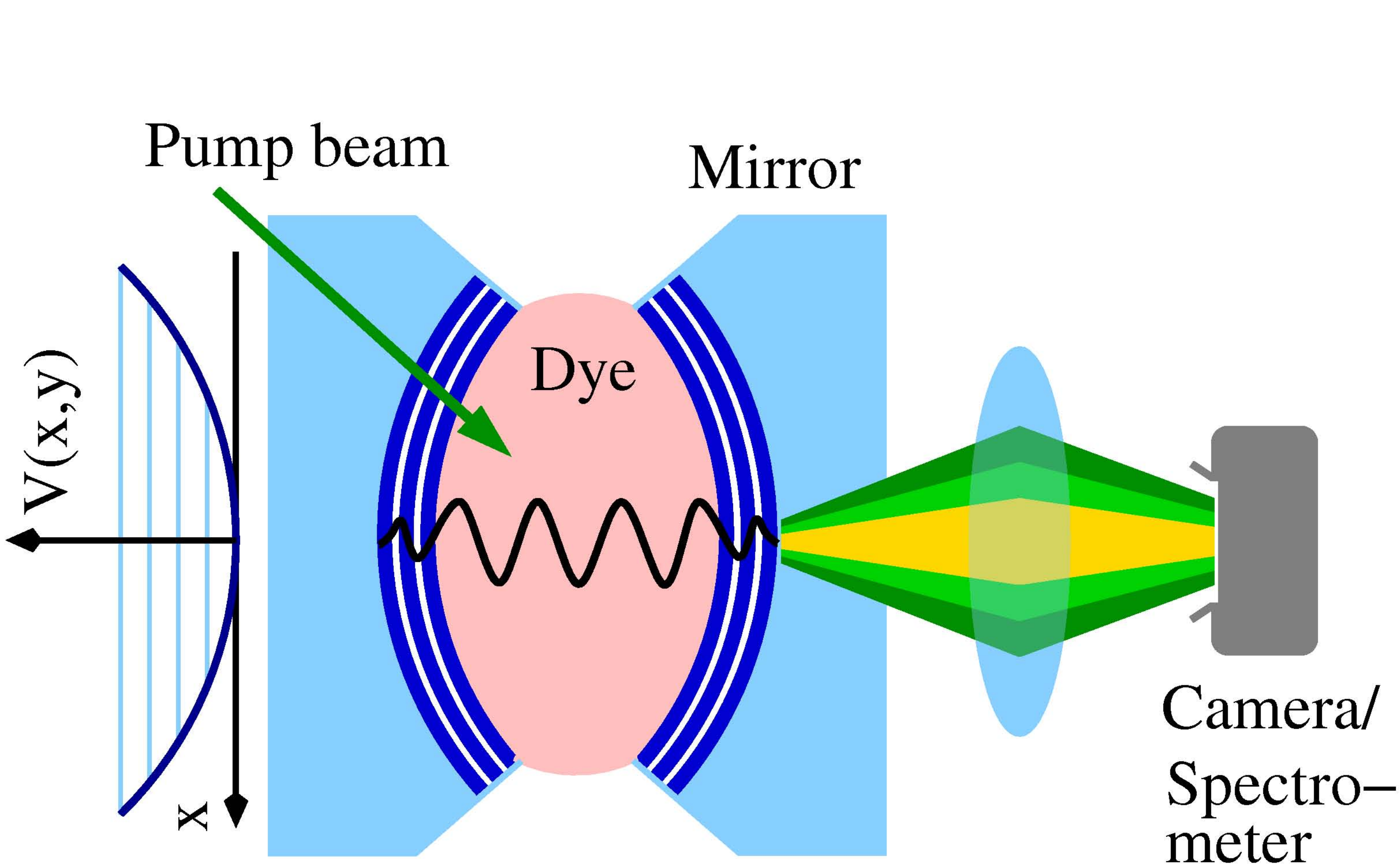}
\caption{\label{fig:figure1}Experimental scheme. (a) Resonant
modes of the microcavity (top) and relative absorption
$\alpha(\nu$) as well as fluorescence strength $f(\nu)$ for
rhodamine 6G (bottom). The height of the bars indicates the
degeneracy of the photon energy. Inside the microcavity, to a good
approximation only fluorescence photons with a longitudinal wave
number of $q=7$ (black bars) exist, which causes this degree of
freedom to be frozen out making the photon gas effectively
two-dimensional. (b) Schematic set-up of the dye-filled
microcavity. The curvature of the mirrors provides a trapping
potential in the transversal plane, as indicated on the left.
Photons with high transversal excitation numbers (green) are
emitted at a larger angle to the optical axis, whereas
transversally less excited radiation (yellow) exhibits a lower
spatial divergence.}

\end{figure}

In our experiment we use a bispherical microresonator filled with
a drop of a dye solution, see Fig.~\ref{fig:figure1}. By multiple
fluorescence and reabsorption the photon gas gets in thermal
contact with the dye solution, and takes over its temperature,
which in our case is room temperature This thermalization process
of the photon gas stems from a thermalization process of the
rovibronic dye state. Due to frequent collisions, on a femtosecond
timescale at room temperature, with the molecules of the solvent
the rovibronic dye state is permanently altered. These processes
are fast enough that both absorption and emission of photons
always take place from highly equilibrated rovibronic dye states.
As a consequence the spectral distributions of absorption
$\alpha(\omega)$ and fluorescence $f(\omega)$ are linked by the
Boltzmann factor
$f(\omega)/\alpha(\omega)\propto\omega^{3}\exp\left(-\hbar\omega/k_{\textrm{B}}T\right)$.
This relation is known as the Kennard-Stepanov relation
\cite{Kennard:1918p1291,Kennard:1927p1292,Stepanov:1957,Stepanov:1957_2}.
Here $\alpha(\omega)$ is the absorption coefficient of the dye
solution, and $f(\omega)$ is the average spectral energy density
of an emitted photon in free space. By multiple absorption and
emission processes this Boltzmann factor is transferred to the
spectral distribution of the photon gas. This can be seen by
investigating the corresponding rate equations for absorption and
emission \cite{Klaers:Thesis2011}. More precisely the photon gas
thermalizes at the spectral temperature defined by
\begin{equation}
T_{\textrm{spec}}(\omega)=\frac{\hbar}{k_{\textrm{B}}}\left(\frac{\partial\ln}{\partial\omega}\frac{\tilde{\alpha}(\omega)}{\tilde{f}(\omega)}\right)^{-1},
\end{equation}
where dimensionless profiles,
$\tilde{\alpha(}\omega)=\alpha(\omega)/\alpha(\omega_{0})$ and
$\tilde{f(}\omega)=f(\omega)\omega^{-3}/f(\omega_{0})\omega_{0}^{-3}$
for an arbitrary $\omega_{0}$, have been used \cite{Klaers:2010}.
However, this spectral temperature coincides for a variety of dye
molecules with the thermodynamic temperature of the dye,
$T_{\textrm{spec}}(\omega)\simeq T$, to good approximation. This
is especially true for the dyes used in the experiment, rhodamine
6G and perylene diimide. Regarding the photon number, it is clear
that the fluorescence induced thermalization process conserves the
photon number on average. A dye molecule can only emit an optical
photon from its electronically excited state if it has absorbed a
photon before. A purely thermal excitation is largely suppressed
at room temperature by a factor
$\exp[-\hbar\omega_\text{c}/k_\text{B} T]\simeq 10^{-37}$ with
$\hbar\omega_\text{c}\simeq 2.1$ eV. In contrast to a black body
radiator, the number of (optical) photons in the cavity is not
determined by the temperature of the dye solution. The chemical
potential of the photon gas thus in general is non-zero. In the
experiment, photons are injected into the cavity by means of an
external pumping laser, whose power thus determines the chemical potential.

The mirror separation is in the micrometer regime, namely
$D_{0}\simeq1.46\,\textrm{\textmu m}$ on the optical axis, and is
of the order of the optical wavelength. Consequently, the free
spectral range of the resonator, of order
$\simeq100\,\textrm{nm}$, becomes comparable to the spectral width
of the dye fluorescence. The presence of the mirrors modifies the
spontaneous emission from the excited molecules inside the cavity
such that predominantly photons of a certain longitudinal mode
number $q$ are found to be emitted, typically $q=7$ in our case.
By this, one degree of freedom is frozen out and the photon gas
becomes two-dimensional, as only the two transversal mode numbers
are variable. Moreover, this establishes a ground state for the
photons with non-vanishing energy \nobreakdash- the transversal
ground mode $\textrm{TEM}_{q00}$. In thermal equilibrium, the
transversal excitations are expected to be Bose-Einstein
distributed. Consequently, for increasing temperature of the dye
solution, the average angle of propagation of the photons with
respect to the optical axis will increase, while for low
temperatures the photons are expected to propagate nearly parallel
to the optical axis.

\subsection{Cavity photon dispersion}

Formally, the photon gas can be mapped onto a two-dimensional,
ideal Bose gas in a harmonic trapping potential. This can be seen
by investigating the energy-momentum relation. The energy of a
photon as a function of longitudinal wave number $k_{z}$ and
transversal wavenumber $k_{r}=\sqrt{k_{x}^{2}+k_{y}^{2}}$ is given
by
\begin{equation} E_{\textrm{ph}}=\frac{\hbar
c}{n}\sqrt{k_{z}^{2}+k_{r}^{2}},
\end{equation}
where $n$ is the index of refraction of the medium. Due to the
mirror curvature the boundary conditions for the photon modes (in
$z$-direction) depend on the distance from the optical axis
$r=|\vec{r}|$. For simplicity these boundary conditions are taken
to be metallic. For the longitudinal wavenumber $k_{z}=k_{z}(r)$
one can make the ansatz \begin{equation}
k_{z}(r)=q\pi/D(r),\end{equation} where\begin{equation}
D(r)=D_{0}-2(R-\sqrt{R^{2}-r^{2}})\end{equation} is the mirror
separation at a distance $r$ from the optical axis and $R$ is the
radius of curvature of the mirrors. To incorporate a possible
nonlinear photon interaction we moreover allow the index of
refraction of the medium to vary in space
$n=n(\vec{r})=n_{0}+\Delta n_{r}$, with $\Delta
n_{r}=n_{2}I(\vec{r})$, in the presence of the light field. Here
$n_{0}$ and $n_{2}$ are the linear and nonlinear indices of
refraction of the medium, respectively, and $I(\vec{r})$ is the
intensity. A non-zero value for $n_{2}$ can arise e.g.\ from a
Kerr-nonlinearity or from thermo-optical effects. With that we
derive\begin{eqnarray}
E_{\textrm{ph}}(\vec{r},k_{r}) & = & \frac{\hbar c}{n(\vec{r})}\sqrt{k_{z}^{2}(r)+k_{r}^{2}}\nonumber \\
 & \hspace{-20mm}= & \hspace{-10mm}\frac{\hbar c}{n_{0}+\Delta n_{r}}\sqrt{\left(\frac{q\pi}{D(r)}\right)^{2}+k_{r}^{2}}\nonumber \\
\nonumber \\ & \hspace{-20mm}\simeq & \hspace{-10mm}\frac{\pi\hbar
cq}{n_{0}D_{0}}+\frac{\hbar cD_{0}}{2\pi
n_{0}q}k_{r}^{2}+\frac{\pi\hbar
cq}{n_{0}D_{0}^{2}R}r^{2}-\frac{\pi\hbar cq}{n_{0}^{2}D_{0}}\Delta
n_{r},\label{eq:Dispersion1}\end{eqnarray} where in the last step
both the paraxial approximation ($r\ll R$, $k_{r}\ll k_{z}$) and
$\Delta n_{r}\ll n_{0}$ has been applied. For an effective photon
mass defined by
\begin{equation}
m_{\textrm{ph}}=\frac{\pi\hbar n_{0}q}{cD_{0}}=\frac{\hbar
n_{0}}{c}k_{z}(0),\label{eq:Masse}
\end{equation}
and a trapping
frequency
\begin{equation}
\Omega=\frac{c}{n_{0}\sqrt{D_{0}R/2}},\label{eq:Fallenfrequenz}
\end{equation}
equation \eqref{eq:Dispersion1} can be written as
\begin{equation}
E_{\textrm{ph}}(\vec{r},k_{r})\simeq\frac{m_{\textrm{ph}}c^{2}}{n_{0}^{2}}+\frac{(\hbar
k_{r})^{2}}{2m_{\textrm{ph}}}+\frac{m_{\textrm{ph}}\Omega^{2}}{2}r^{2}-\frac{m_{\textrm{ph}}c^{2}}{n_{0}^{3}}n_{2}I(\vec{r}),\label{eq:E_ph}
\end{equation}
When the interaction term is omitted, i.e.\ for $n_{2}=0$, this is
the energy-momentum relation of a non-rel\-a\-tiv\-is\-tic,
massive particle in two dimensions, bound by a harmonic trapping
potential. The standard quantization procedure then leads to the
well known energy spectrum of the bispherical resonator
\cite{Kogelnik:1966p1652}, with the energy levels equidistantly
separated by $\hbar\Omega$ and the degeneracy increasing linearly.
After rescaling the energy such that the lowest harmonic
oscillator mode has zero energy, and defining the transversal
photon energy $u=E_{\textrm{ph}}-m_{\textrm{ph}}c^{2}/n_{0}^{2}-\hbar\Omega$
the degeneracy can be written as
\begin{equation}
g(u)=2(u/\hbar\Omega+1).\label{eq:degeneracy}\end{equation} The
prefactor $2$ here goes back to the two-fold polarization
degeneracy.

%%%%%%%%%%%%%%%%%%%%%%%%%%%%%%%%%%%%%%%%%%%%%%%%%%%%%%%%%%%%%%%%%
\subsection{Statistical theory of the photon gas}
%%%%%%%%%%%%%%%%%%%%%%%%%%%%%%%%%%%%%%%%%%%%%%%%%%%%%%%%%%%%%%%%%
The thermodynamics of such a system is well known: a harmonically
confined two-dimensional Bose gas shows a phase transition to a
BEC at low temperatures or high particle numbers
\cite{Bagnato:1991p552,Petrov:JPhysIV2004,Mullin:1997p1821,Mullin:1998p1818}.
The average occupation number of a certain transversal energy $u$
is given by the Bose-Einstein distribution\begin{equation}
n_{T,\mu}(u)=\frac{g(u)}{e^{\frac{u-\mu}{k_{\textrm{B}}T}}-1},\label{eq:BEdistr}\end{equation}
where the chemical potential of the photons is implicitly defined
by\begin{equation}
N_{\textrm{ph}}=\sum_{u}n_{T,\mu}(u),\label{eq:N(mu)}\end{equation}
with $N_{\textrm{ph}}$ as the total (average) photon number. With
that, the number of photons in the transversally excited resonator
states is seen to saturate at a critical particle number
$N_{\textrm{c}}$ given by
\cite{Bagnato:1991p552,Petrov:JPhysIV2004,Mullin:1997p1821,Mullin:1998p1818}
\begin{equation}
N_{\textrm{c}}\simeq\frac{\pi^{2}}{3}\left(\frac{k_{\textrm{B}}T}{\hbar\Omega}\right)^{2}.\end{equation}
If the total photon number is increased beyond this bound, a
macroscopic fraction is expected to condense into the transversal
ground state. In our experiments, the trapping frequency is of
order $\Omega\simeq10^{11}\,\textrm{Hz}$, which is roughly 9
orders of magnitude higher than for experiments with ultracold
atomic gases. At room temperature, $T=300\,\textrm{K}$, this gives
a critical photon number of magnitude $N_{\text{c}}\simeq10^{5}$,
which is experimentally feasible. In principle, it would also be
possible to lower the temperature of the dye solution below the
critical temperature,
\begin{equation}
T_{\textrm{c}}\simeq\frac{\sqrt{3}\hbar\Omega}{\pi
k_{\textrm{B}}}\sqrt{N_\textrm{ph}}=\frac{\sqrt{6}\hbar c}{\pi
k_{\textrm{B}}n_{0}}\sqrt{\frac{1}{D_{0}}\frac{N_\textrm{ph}}{R}},\label{eq:Tc}
\end{equation}
to trigger the condensation for a fixed photon number. However,
this is experimentally much more demanding. Strictly speaking,
phase transitions can only occur in the thermodynamic limit. From
eq.~\ref{eq:Tc} we see that this thermodynamic limit has to be
$N_{\textrm{ph}}\rightarrow\infty$, $R\rightarrow\infty$ with
$N_{\textrm{ph}}/R=\textrm{const}$, as only this retains
$T_{\textrm{c}}$. A limit invoking $D_{0}\rightarrow\infty$ is
unphysical because the two-dimensionality of the photon gas will
be lost. Other than in the confined case considered here, there is
no thermodynamic limit for a macroscopic ground state occupation
at finite temperatures for the homogeneous two-dimensional Bose
gas
\cite{Bagnato:1991p552,Petrov:JPhysIV2004,Mullin:1997p1821,Mullin:1998p1818}.
Correspondingly, such a system is always multimode in the
thermodynamic limit.

In contrast to polariton experiments
\cite{Kasprzak:2006p444,Balili:2007p1342,Kasprzak:2008p1293,Kasprzak:2008p716,Lagoudakis:NaturePhysics2008,Amo:NaturePhysics2009},
no strong atom-light coupling takes place in our experiment. The
frequent collisions with the molecules of the solvent leads to a
rapid decoherence of the dye state, i.e. the decoherence rate is
bigger than the Rabi frequency. This prevents a coherent energy
exchange between photonic and molecular excitations
\cite{Yokoyama:1989p2104,DeAngelis:2000p484}. Consequently, the
eigenstates of the system are best described by the uncoupled
photonic and molecular degrees of freedom. At any given point in
time a certain number of photons, $N_\textrm{ph}$, and excited dye
molecules, $N_\textrm{exc}$, populate the cavity. In a stationary
state, i.e.~no net flux, the ratio of these numbers will be linked
by the ratio of the lifetimes of both species,
$N_{\textrm{exc}}/N_{\textrm{ph}}\simeq\tau_{\textrm{exc}}/\tau_{\textrm{ph}}$.
The lifetime of a photon $\tau_{\textrm{ph}}$ here denotes the
average time between emission and reabsorption in the medium. For
typical experimental parameters $\tau_{\textrm{ph}}$ is of
magnitude $10\,\textrm{ps}$. The lifetime of the excited dye
molecules $\tau_{\textrm{exc}}$ is well known to get modified by
the presence of the cavity with respect to its free space value
\cite{Drexhage:1974,Hulet:1985p554,JHE:1987p1903,DEMARTINI:1987p658,DEMARTINI:1991p2107,MILONNI:2007p655,Loudon:2007p2116,Meschede:PhysicsReports1992,Walther:Reports2006}.
Its exact value has a complicated dependence on the detailed
experimental conditions. Moreover, it will be influenced by the
presence of other photons due to stimulated emission.
Nevertheless, we expect it to be of order 1\,ns for the used
optical powers, and thus it remains clearly bigger than the photon
lifetime under typical experimental conditions
\cite{DEMARTINI:1987p658,DEMARTINI:1991p2107}. As a result, the
number of excited molecules will also be much bigger than the
number of photons. This situation is best described by a
grandcanonical particle exchange, where the photon gas exchanges
particles with the reservoir of excited dye molecules. The
question regarding the suitable statistical ensemble is important
because the fluctuations of the photon number \nobreakdash- and
with that the coherence properties of the light \nobreakdash- will
strongly depend on the statistical ensemble, even in the
thermodynamical limit
\cite{Fujiwara:1970p993,ZIFF:1977p513,Kocharovsky:2006p985}.

\subsection{Rate equation model}

Semiclassical laser theory builds upon rate equation models, where
laser action is predicted when the gain from stimulated emission
exceeds the total loss \cite{ScullyZubairy}. In an analogous way
we use a multimode theory based on rate equations to describe the
thermalization in the here investigated dye-filled optical
microcavity. The dye molecule is modelled by a Jablonski energy
level scheme \cite{Lakowicz:1999} commonly used in the
Kennard-Stepanov theory of absorption and emission
\cite{Kennard:1918p1291,Kennard:1927p1292,Stepanov:1957,Stepanov:1957_2,Sawicki:1996p2109}.
The present calculation neglects cavity specific modifications of the light-matter interaction, as well as fluctuations of the particle number, both issues that will be incorporated in a
subsequent publication \cite{janklaerstobepublished}.

\begin{figure}
\noindent \begin{centering}
\includegraphics[width=7cm]{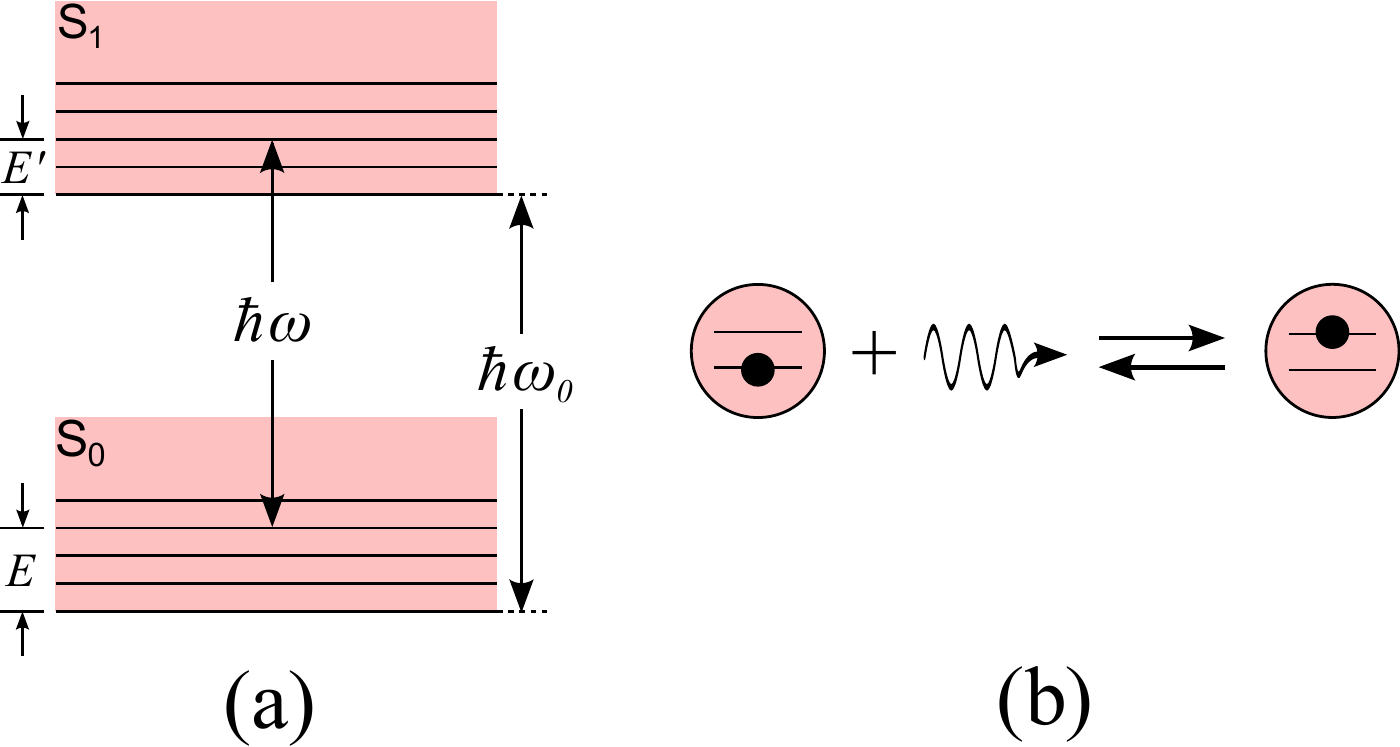}
\par\end{centering}

\caption{\label{fig:figure_rate_eqs}(a) Energy level scheme of a
dye molecule with an electronic ground state $S_0$ and an excited
state $S_1$ respectively, where "S" denotes singlet, which are
separated by an energy $\hbar\omega_0$. Photons of a frequency
$\omega$ are emitted and absorbed from within the rovibrational
substates $E$ and $E'$ respectively. (b) Chemical reaction type
exchange between different particle species, namely ground state
dye molecules, photons and excited dye molecules.}

\end{figure}

Consider a system with one ground and one excited electronic
state, each with a manifold of rovibrational substates, see
Fig.~\ref{fig:figure_rate_eqs}a. Let $\hbar\omega_0$ denote the
energy splitting between the lowest rovibronic states (zero-phonon line). Furthermore, $E$ and $E'$ are considered to be rovibrational energies of ground and excited state and $\varrho(E)$,  $\varrho'(E')$ denote the corresponding rovibronic densities of states. The distribution functions $p(E)$, as
well as $p'(E')$, are proportional to the probability of finding a rovibronic state with energy $E$ and $E'$; for example $p(E)=\exp(-E/k_{\text{B}}T)$.
With this, we expect the following total rates for spontaneous emission,
stimulated emission and absorption of a photon at frequency
$\omega$:
\begin{subequations}
\begin{align}
R_{\text{spon}}(\omega) &= \int \text{d}E' \: n_e(E') \: A\,\,(E',\omega) \;\;, \label{eq:rate_spont} \\
R_{\text{stim}}(\omega) &= \int \text{d}E' \: n_e(E') \: B'(E',\omega) \: u(\omega) \;\;, \label{eq:rate_stim}\\
R_{\text{abs}}(\omega)\  &= \int \text{d}E\, \:\, n_g(E\,) \:
B\,\,(E\,\,,\omega) \: u(\omega) \;\;, \label{eq:rate_abs}
\end{align}
\end{subequations}
where $u(\omega)$ is the spectral energy density per unit volume
of the radiation field, $A$, $B$ and $B'$ are the corresponding
Einstein coefficients, and the occupation densities $n_e(E')$ and
$n_g(E)$ of the excited and ground state populations are given by
\begin{subequations}
\begin{align}
n_e(E') &= N_e \: \frac{\varrho'(E') \: p'(E')}{W'} \label{eq:na} \;\: ,\\
n_g(E\,) &= N_g \: \frac{\varrho\,(E\,\,) \: p\,(E\,\,)}{W\,\,}
\;\; . \label{eq:ng}
\end{align}
\end{subequations}
Here $N_g$ and $N_e$ denote the number of molecules in ground and
excited states in the optically active volume. Further, $W$ and
$W'$ are statistical weights given by
$W=\int\text{d}E\:\varrho\,(E)p(E)$ and
$W'=\int\text{d}E'\:\varrho'\,(E')p'(E')$.  This treatment assumes
that cavity mirror losses and the coupling to unconfined optical
modes are small, i.e. the rates associated with the corresponding
loss processes are small compared to the processes accounted for
in \eqref{eq:rate_spont}--\eqref{eq:rate_abs}. Similarly, the rate
for pumping, which compensates for these losses, will then be
small. These conditions will be valid if a photon collides several
times with dye molecules in the cavity before being lost, as can
be achieved by reabsorption, i.e.\ by recapturing fluorescence.

The Einstein coefficients are not independent, their connection is
given by
\begin{equation}
A(E',\omega)\:\text{d}E' = \hbar\omega D(\omega)
B'(E',\omega)\:\text{d}E',\label{eq:AB}
\end{equation}
(A-B relation), with $D(\omega)$ as the spectral mode density per
unit volume, and by
\begin{equation}
\varrho'(E') B'(E',\omega)\:\text{d}E'= \varrho(E)
B(E,\omega)\:\text{d}E,\label{eq:BB}
\end{equation}
which follows from a detailed balance argument
\cite{Sawicki:1996p2109}. If we assume that the rovibrational dye
state is in thermal equilibrium in both the electronic ground and
excited states manifold respectively, which for many dyes is a
reasonable assumption given that the typical time for a collision
with solvent molecules is in the femtosecond regime at room
temperature, while the electronic lifetimes are typically of the
order of a nanosecond \cite{Lakowicz:1999}, we can apply Boltzmann
distributions for $p(E)$ and $p'(E')$.

From the rate equations \eqref{eq:rate_spont}--\eqref{eq:rate_abs}
the net variation of the photon number $n(\omega)$ at frequency
$\omega$ is
\begin{subequations}
\begin{align}
\frac{\text{d}n(\omega)}{\text{d}t} &= R_{\text{spon}}(\omega) + R_{\text{stim}}(\omega) - R_{\text{abs}}(\omega)\\
&=\int \text{d}E\ \varrho(E) e^{-E/k_\textrm{B}T} B(E,\omega) \;\; \times \nonumber\\
&\left[\frac{1}{W'} N_e e^{-\hbar(\omega-\omega_0)/k_\textrm{B}T}\lbrack \hbar\omega D(\omega)+u(\omega)\rbrack\right.  \nonumber\\
&-\left.\frac{1}{W}N_g u(\omega)\right], \label{eq:dn_dt_2}
\end{align}
\end{subequations}
where in the second step we have used the energy conservation
relation $\hbar\omega + E = \hbar\omega_0 + E'$ and the equations
\eqref{eq:AB} and \eqref{eq:BB} to combine all integrals.

In a steady state, $\text{d}n(\omega)/\text{d}t=0$, the term in
squared brackets must vanish. This yields the equilibrium photon
energy distribution
\begin{equation}
n(\omega)=\frac{g(\omega)}{\frac{N_g W'}{N_e W}\
e^{\hbar(\omega-\omega_0)/k_\textrm{B}T}-1},
\label{eq:photon_number_distribution}
\end{equation}
where additionally $n(\omega) = \int \text{d}V\
\frac{u(\omega)}{\hbar\omega}$ and $\int \text{d}V\
D(\omega)=g(\omega)$, see eq.~\eqref{eq:degeneracy}, were
employed. In chemical equilibrium, the chemical potentials of all
species are linked by $\mu_{\textrm{ph}}+\mu_g=\mu_e$
(Fig.~\ref{fig:figure_rate_eqs}b), where $\mu_\textrm{ph}$ is the
chemical potential of the photons and $\mu_e$, $\mu_g$ are the
chemical potentials of excited and ground state dye molecules,
respectively. From this one can easily show
\cite{janklaerstobepublished} that the fugacity of the photons is
determined by the excitation level in the dye medium
\begin{equation}
e^{\mu_\textrm{ph}/k_\textrm{B} T}=\frac{N_e W}{N_g W'}\
e^{\hbar\omega_0/k_\textrm{B} T} \;\; ,
\label{eq:chemical_potential}
\end{equation}
or, if we use a renormalized chemical potential
$\mu=\mu_\textrm{ph}-\hbar\omega_\text{c}$,
\begin{equation}
e^{\mu/k_\textrm{B} T}=\frac{N_e W}{N_g W'}\
e^{\hbar(\omega_0-\omega_\text{c})/k_\textrm{B} T} \;\;.
\label{eq:chemical_potential_2}
\end{equation}
Thus, in accordance with the usual convention, the photon number
approaches infinity, $N_\text{ph}\rightarrow\infty$, for a
vanishing chemical potential, $\mu \rightarrow 0$ with $T=\text{const}$. When we combine
the equations \eqref{eq:photon_number_distribution} and
\eqref{eq:chemical_potential_2}, we arrive at the Bose-Einstein
distribution, eq.~\eqref{eq:BEdistr}.

As it is well known from laser theory \cite{ScullyZubairy}, the
rate equations \eqref{eq:rate_spont}--\eqref{eq:rate_abs} do not
allow for an inversion of a specific transition, and one therefore
has to include a pumping term to the equations to model lasing.
However, the rate equations do predict thermally distributed
photon energies for certain conditions, as we have shown. This
includes that the statistical weight of the resonator ground mode
can become macroscopic. Put differently, 'single mode'-behaviour
can arise even in thermal equilibrium and without inversion. This
point is however only revealed in a statistical multi-mode
treatment and is missed in the usual laser theory. That the photon
condensate can arise in a non-inverted medium can also be
understood by the capability of our system to reuse spontaneous
emission. In a macroscopic laser reabsorption nearly inevitably
leads to photon loss due to spontaneous emission into non-confined
modes. However, if spontaneous emission is directed into the
cavity, where it gets reabsorbed, the excitation will not be lost.
Therefore spontaneous emission in general is not a loss channel in
our system. Recapturing of fluorescence (or photon recycling) is a
mechanism also known for "zero threshold" microlasers
\cite{DeMartini:1988p632,Yokoyama:1992p2123,Yamamoto:1992p2125}.
In those works one tries to capture spontaneous emission into a
single cavity mode, while in the present scheme it is sufficient
if the spontaneous photons are directed into one of the (many)
transversal modes. The thermalization process will select the
resonator ground state to have the biggest statistical weight,
which can even be macroscopic, when condensation sets in.

\section{Experimental setup\label{sec:ExperimentalSetup}}

\begin{figure*}
\noindent \begin{centering}
\includegraphics[width=10cm]{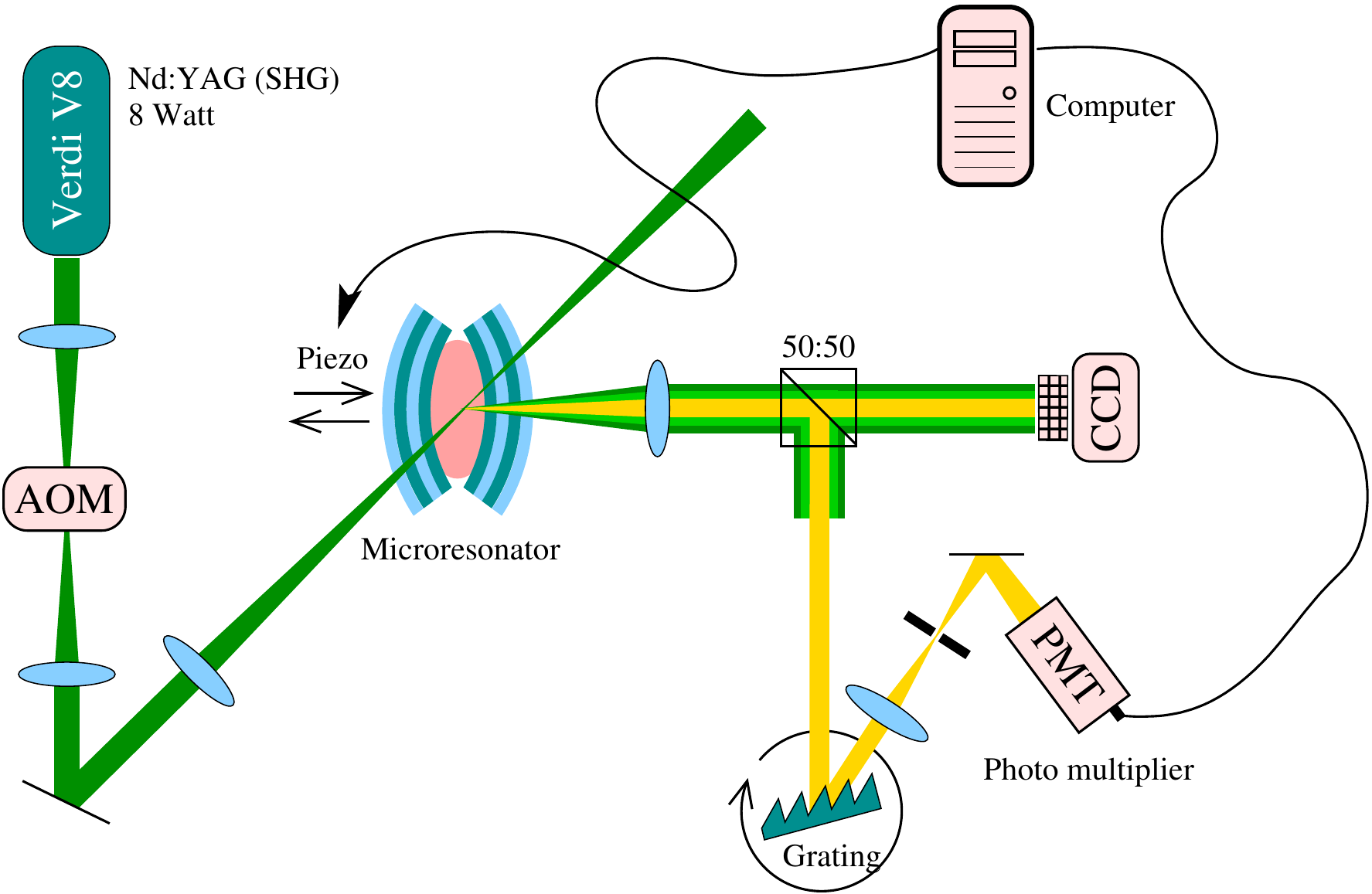}
\par\end{centering}

\caption{\label{fig:figure2}Schematic set-up of the microcavity
experiment. The microresonator is pumped at an angle of
45\textdegree{} to the optical axis. The radiation emerging from
the microresonator is examined both spatially and spectrally. }

\end{figure*}

The experimental setup, as shown in Fig.~\ref{fig:figure2},
consists of a microresonator, a pump source and an analyzing
section. The microresonator is formed by two highly reflecting,
spherically curved, dielectric mirrors which are typically used
for cavity ring-down spectroscopy. Their reflectivity is bigger
than $0.99997$ for the wavelength region $(500-590)\,\textrm{nm}$,
with a reflectivity maximum of $0.999985$ at
$\lambda\approx535\,\textrm{nm}$. The latter corresponds to a
finesse of $F\approx200000$ for the empty cavity. Two different
radii of curvatures are available, $R=1\,\textrm{m}$ and
$R=6\,\textrm{m}$, respectively. To allow for a mirror separation
of a few halfwaves on the optical axis despite the curvature, one
of the mirrors is cut to a surface area of
$1\,\textrm{mm}\times1\,\textrm{mm}$. The edges are carefully
polished to minimize possible stress from contact points between
the mirrors. To verify that the mirror's reflectivity remains
unaffected by the cutting and subsequent polishing procedure
cavity ring-down measurements have been performed.

The microresonator is pumped at an angle of 45\textdegree{} with
respect to the optical axis. At this angle and for a certain
polarisation the mirrors show their first reflectivity minimum
allowing a pump beam transmission of over 80\%. To pump the
resonator in this way is advantageous because resonance conditions
and mode matching do not have to be considered. The pump beam can
be placed at an arbitrary position inside the resonator plane
independent of the cavity resonances. One of the mirrors is
mounted on a translation stage with a piezoelectric actuator to
allow fine tuning of the mirror separation, which can be
computer-controlled to achieve an electronic stabilization. For
that the fluorescence spectra emitted from the cavity are captured
by a spectrometer and are used as an input for the stabilization
protocol that controls the voltage driving the piezoelectric
actuator. The bandwidth of this control loop is sufficient to
compensate thermally induced fluctuations of the cavity length.
Faster fluctuations are reduced by mechanically decoupling the
cavity from the optical table and by the damping caused by the dye
solution between the mirrors.

\begin{figure}
\noindent \begin{centering}
\includegraphics[width=0.5\textwidth]{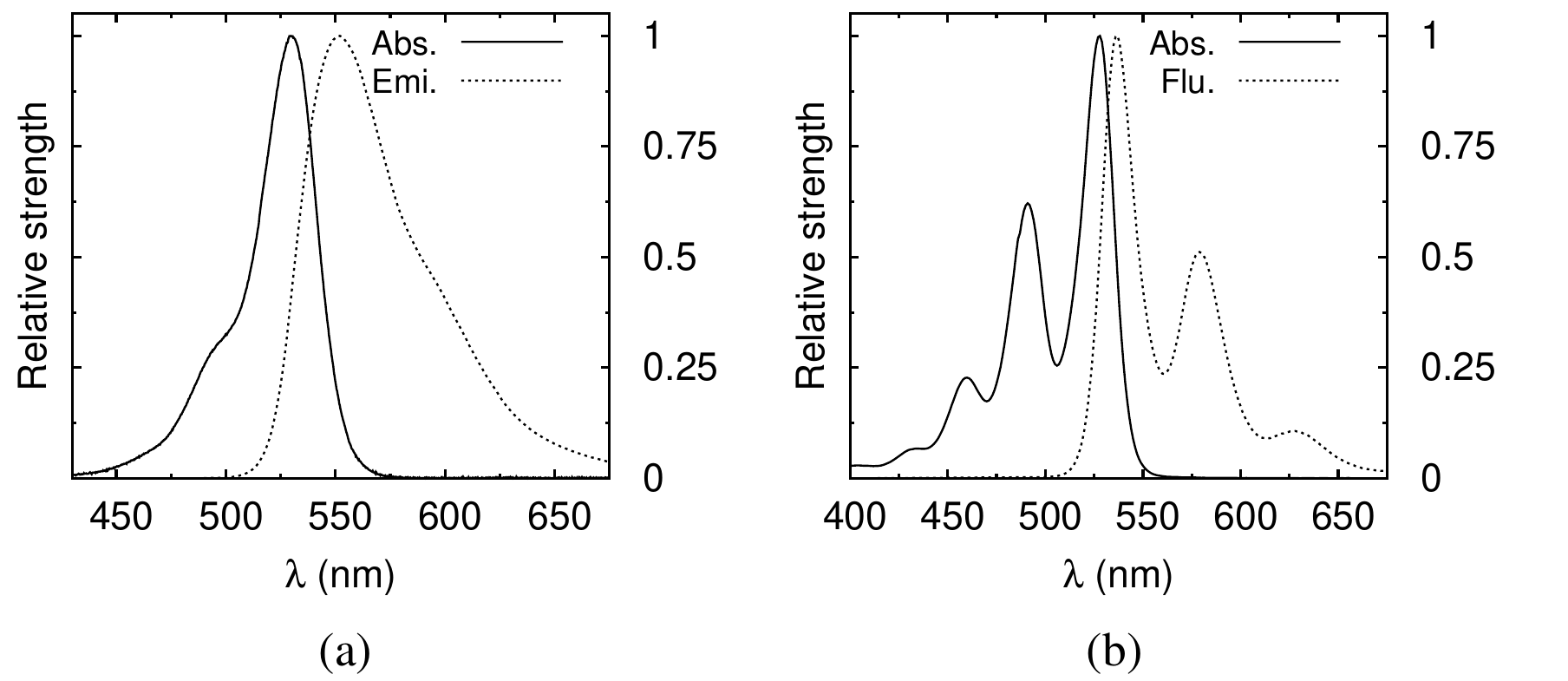}
\par\end{centering}

\caption{\label{fig:figure3}Relative strengths of absorption
$\alpha(\lambda)/\alpha_{\textrm{max}}$ and fluorescence
$f(\lambda)/f_{\textrm{max}}$ for the dyes (a) rhodamine 6G and
(b) perylene diimide.}

\end{figure}

The used dyes are either rhodamine 6G or perylene-diimid (PDI),
whose absorption and fluorescence spectra are shown in
Fig.~\ref{fig:figure3}. The dyes show a quantum efficiency close
to unity in the used spectral region,
$\Phi_{\textrm{R6G}}\approx0.95$ \cite{Magde:2002p2101} and
$\Phi_{\textrm{PDI}}\approx0.97$ \cite{Wilson:2009p2118}, and have
spectral temperatures close to the thermodynamic temperature of
the surrounding solvent, $T_{\textrm{spec}}(\omega)\simeq T$. For
rhodamine we either use methanol or ethylene glycol as solvent,
for PDI aceton is used. Repeated filtering of the solutions is
necessary to remove unsolved dye particles and other
contaminations. This filtering has proven to be crucial for the
reliability of the experiment. To achieve reabsorption of the
microcavity radiation fairly high concentrated dye solutions have
to be employed. In principle high dye concentrations can lead to
radiationless deactivation. However, the concentration used here,
$1.5\times10^{-3}\,\textrm{Mol/l}$ for rhodamine 6G, is still one
order of magnitude lower than the concentration where fluorescence
quenching notably sets in
\cite{Penzkofer:1986p1323,Penzkofer:1987p1311,Fischer:1996p1357}.
As pump source we use a frequency doubled Nd:YAG laser at a
wavelength of $532\,\textrm{nm}$ which is switched by two
acousto-optic modulators operating in series to ensure a high
on/off contrast. The analysis is performed by a spectrometer and a
camera. For that the emitted radiation is split by a
non-polarizing beam splitter. One part of the beam is mapped by an
imaging system onto the chip of a color CCD camera, where a
magnified real image of the photon gas is generated. The other
part of the beam is directed into a spectrometer. We use either a
commercially available one, with optical resolution
$\simeq2\,\textrm{nm}$, or a self-built spectrometer. Care has to
be taken when the spectrometer has an entry slit, as it is the
case for the commercial spectrometer, where the width is
$15\,\textrm{\textmu m}$. Due to their bigger mode diameter and
higher divergence the coupling of higher transversal modes through
the slit typically will be much less efficient than, for example,
for the ground mode $\textrm{TEM}_{\textrm{q}00}$. Consequently,
the entry slit will not be neutral in color anymore. One way to
overcome this problem is to use a diffusing plate in the light
path that cancels out the correlation between position, angle and
color. A more advantageous way is to omit the entry slit
alltogether and to directly pass the light onto a diffractive
grating. This is possible because the emitted radiation from the
microcavity is only weakly divergent and can be sufficiently
collimated before being spectrally decomposed. This scheme is
realized in our self-built spectrometer, using a rotating
diffractive grating with $2400\,\textrm{lines/mm}$ and a
photomultiplier tube as detector. However, both methods described
above deliver largely equivalent spectra.

\section{Thermalization of the transversal photon
state}\label{sec:thermalization}

\subsection{Spectral and spatial photon distribution}

\begin{figure}
\noindent \begin{centering}
\includegraphics[width=8.5cm]{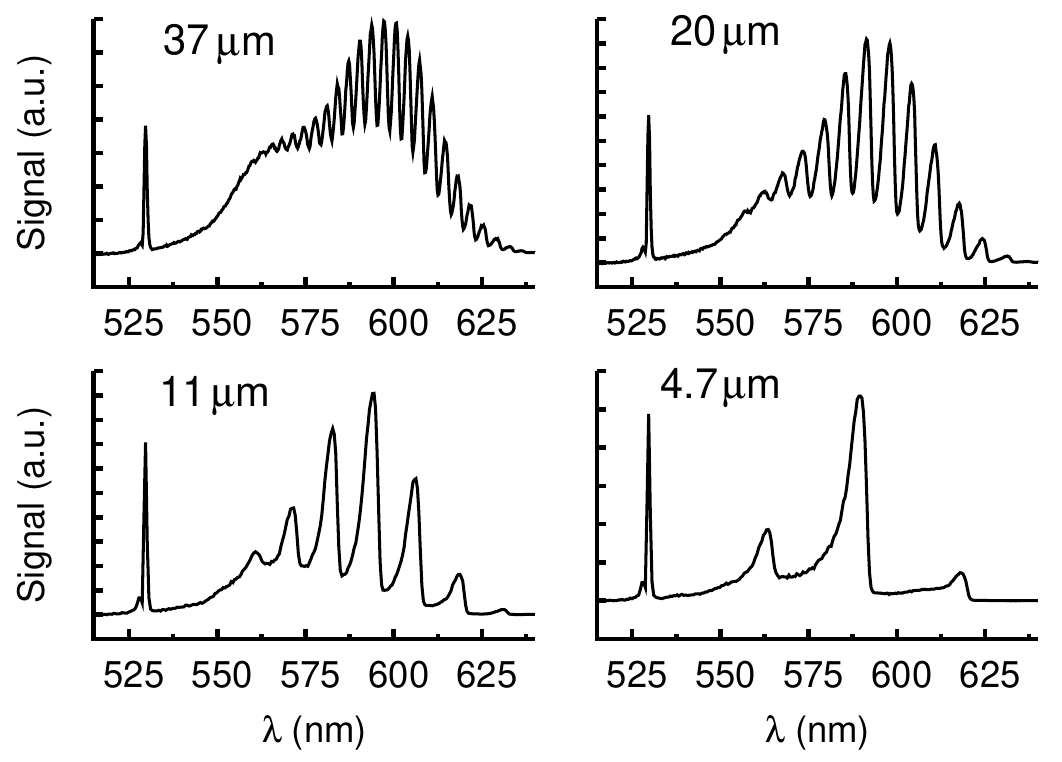}
\par\end{centering}

\caption{\label{fig:figure4}Spectral distribution of the radiation
emitted from the microcavity for different mirror separations,
$D\approx \{37,20,11,4.7\}\,\mu$m, corresponding to different cavity orders $q\approx \{175,95,50,22\}$. (Rhodamine 6G in methanol,
$\varrho=1.5\times10^{-3}\,\textrm{Mol/l}$, mirror curvatures
$R_{1}=1\,\textrm{m}$ and $R_{2}=6\,\textrm{m}$) }

\end{figure}

\begin{figure}
\noindent \begin{centering}
\includegraphics[width=0.40\textwidth]{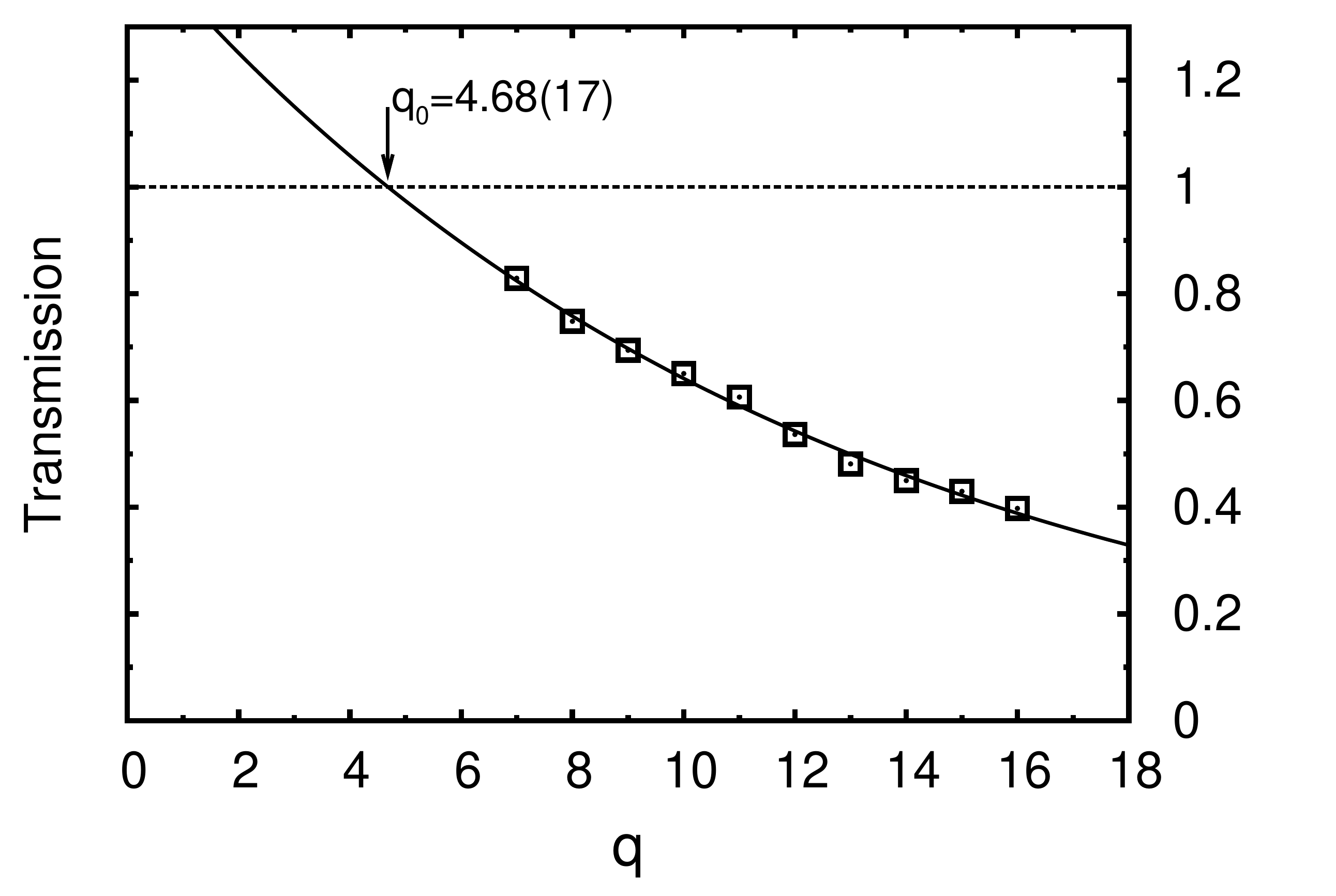}
\par\end{centering}

\caption{\label{fig:figure5}Transmission of the pump beam
(squares) as a function of the longitudinal mode number $q$. The
solid line corresponds to the Lambert-Beer law
$I(q)/I_{0}=\exp(-\alpha(q-q_{0})\lambda_\text{c}/2n_{0})$ with
the fitting parameter $q_{0}=4.68\pm0.17$. This measurement shows
that the effective end nodes of the optical resonance are located
inside the mirrors. (Rhodamine 6G in methanol,
$\varrho\approx2\times10^{-2}\,\textrm{Mol/l}$, mirror curvatures
$R_{1}=R_{2}=1\,\textrm{m}$) }

\end{figure}

Figure~\ref{fig:figure4} shows typical spectral distributions of
the emitted microresonator light for decreasing mirror separation
(from upper left to lower right). By investigating the free
spectral range the longitudinal mode number of the photons can be
determined. For very small mirror separations a regime is reached
where the photon gas gets two-dimensional, i.e. where only photons
of a single longitudinal mode number are emitted. The shortest
mirror separation that can be reached in our experiment
corresponds to $q=7$ halfwaves. The limiting factor that prevents
a further reduction of $q$ is the penetration of the light field
into the mirror material. This can be seen by determining the
transmission coefficient of the pump light, which is illustrated
in Fig.~\ref{fig:figure5}. For this measurement very high
concentrated dye solutions are used (rhodamine 6G at
$2\times10^{-2}\,\textrm{Mol/l}$). As expected, the transmission
of the pump light shows an exponential decay with increasing $q$
(Lambert-Beer law). However, the decay curve extrapolates to full
transmission at a value of $q_{0}=4.68\pm0.17$. Apparently the
thickness $D_{\textrm{dye}}$ of the dye film is not directly
proportional to $q$ but shows an offset, following
$D_{\textrm{dye}}(q)\propto(q-q_{0})$. For example, the dye film
thickness at $q=7$ is just over two halfwaves. The effective end
nodes of the light field are located about $2.3$ halfwaves
(corresponding to $400-450\,\textrm{nm}$) inside the mirrors.
Longitudinal mode numbers $q<5$ are therefore not possible with
the used mirrors.

The true ground state of the resonator is the halfwave resonance
$\textrm{TEM}_{100}$. The freezing out of the longitudinal mode
number at mirror separations corresponding to $q\gtrsim7$ is not
obvious at first sight. Fluorescence into
$\textrm{TEM}_{\textrm{8mn}}$ modes, or into transversally low
excited $\textrm{TEM}_{\textrm{6mn}}$ modes are energetically
excluded due to the finite spectral width of the dye emission. But
fluorescence into transversally high excited
$\textrm{TEM}_{\textrm{6mn}}$ modes is energetically possible.
However, due to their much higher mode volume and consequently
smaller overlap with an emitting point dipole, decays into higher
$\textrm{TEM}_{\textrm{6mn}}$ modes are expected to be much slower
than emission processes into the energetically equivalent but
transversally low excited $\textrm{TEM}_{\textrm{7mn}}$ modes.
This is indeed also seen in the experimental spectra
(Fig.~\ref{fig:figure4}), where a preference for low transversal
modes is already seen for rather large mirror separations. Thus
mode hopping to longitudinally lower excited modes is not found to
be a relevant loss mechanism for the two-dimensional
photon gas. %
\begin{figure}
\noindent \begin{centering}
\includegraphics[width=7cm]{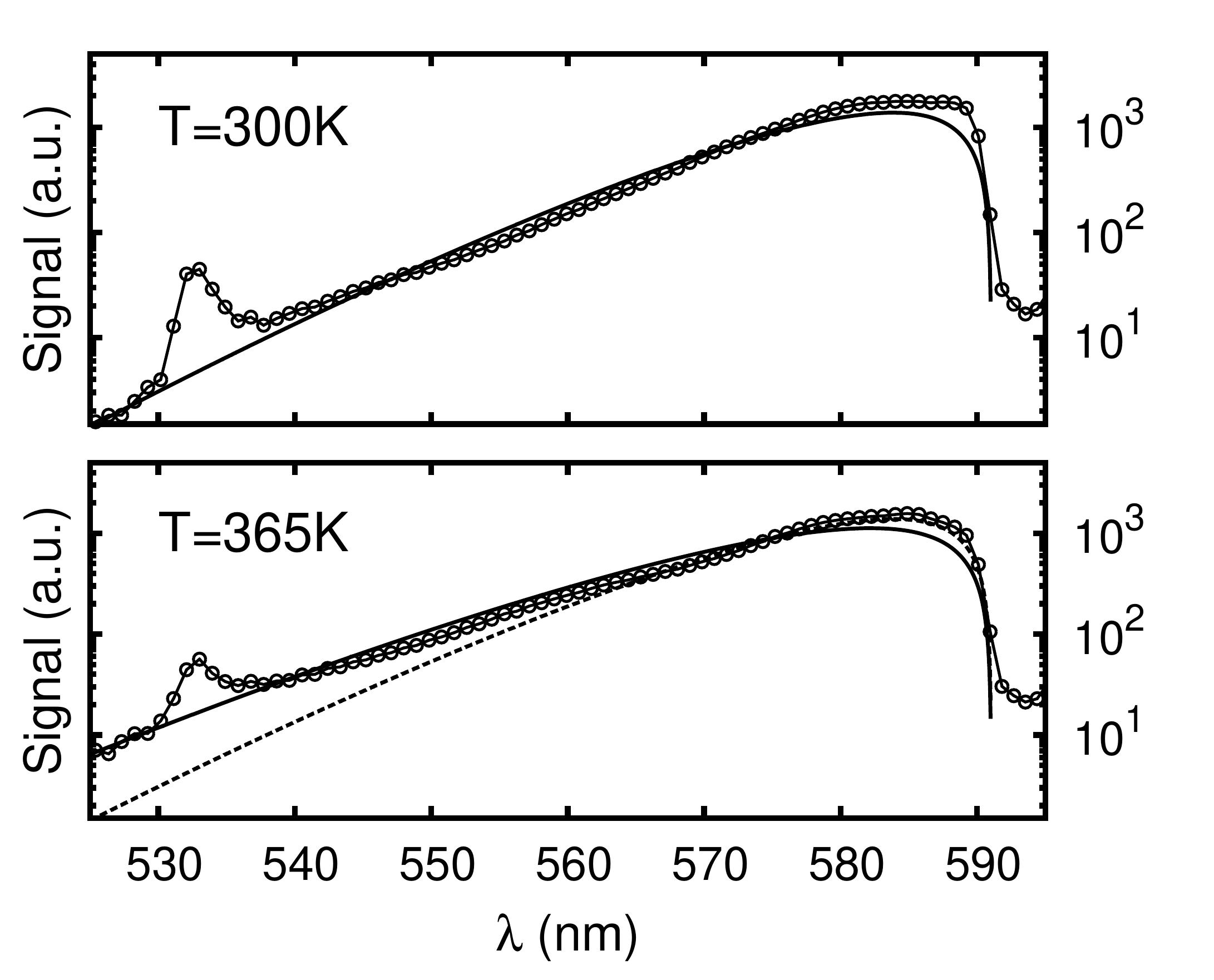}
\par\end{centering}

\caption{\label{fig:figure6}Spectral distribution of the radiation
emitted from the microcavity far below the critical photon number
at $T=300\,\textrm{K}$ und $T=365\,\textrm{K}$ (circles). The
spectra are in good agreement with the Boltzmann-distributed
photon energies (lines). For comparison a $T=300\,\textrm{K}$
Boltzmann-distribution is additionally plotted in the bottom graph
(dashed line). (Rhodamine 6G in ethylene glycol,
$\varrho=5\times10^{-4}\,\textrm{Mol/l}$, mirror curvatures
$R_{1}=R_{2}=1\,\textrm{m}$, $q=7$) }

\end{figure}
\begin{figure}
\noindent \begin{centering}
\includegraphics[width=7cm]{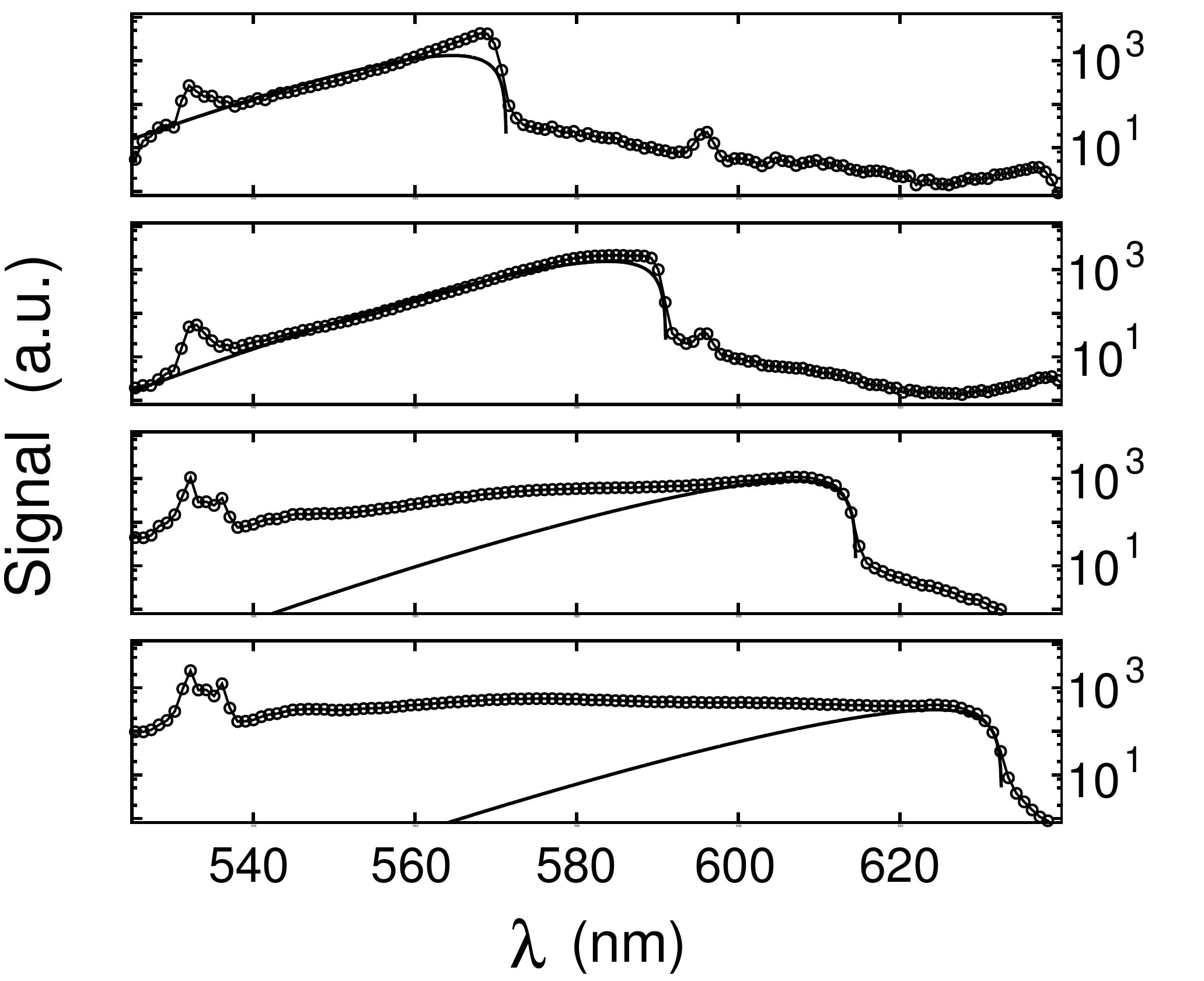}
\par\end{centering}

\caption{\label{fig:figure7}Normalized spectral distribution of
the radiation emitted from the microcavity (circles) for 4
different cut-off wavelengths $\lambda_\text{c}\approx\{570, 590, 615, 625\}\,\textrm{nm}$. In addition,
Boltzmann-distributed photon energies for $T=300\,\textrm{K}$ are
plotted (lines). (Experimental parameters as in
Fig.~\ref{fig:figure6}) }

\end{figure}

Measurements of the spectral photon distribution in the
microresonator at room temperature, $T=300\,\textrm{K}$, and for a
heated resonator setup at $T=365\,\textrm{K}$ can be seen in
Fig.~\ref{fig:figure6}. The power of the transmitted light is
$P_{\textrm{out}}=(50\pm5)\,\textrm{nW}$. From this value an
average photon number of $N=60\pm10$ inside the resonator can be
derived. A numerical solution of eq.~\eqref{eq:N(mu)} then leads
to a chemical potential of $\mu/k_{\textrm{B}}T=-6.76\pm0.17$
($T=300\,\textrm{K}$) and $\mu/k_{\textrm{B}}T=-7.16\pm0.17$
($T=365\,\textrm{K}$). These measurements are thus performed far
from the phase transition that is expected to set in for $\mu$
getting close to zero. Consequently the term '-1' in the
denominator of the Bose-Einstein distribution,
eq.~\eqref{eq:BEdistr}, can be neglected and the spectral
distribution of the photons is expected to be Boltzmann-like.
Besides some deviations around 532\,nm caused by scattered pump
light the measured spectra show a good agreement with the
theoretical expectations both for $T=300\,\textrm{K}$ and for
$T=365\,\textrm{K}$, respectively. This can be interpreted as
evidence for a thermalization process of the transversal photon
state.

As stated before a thermalization process is expected to occur
only if a thermal contact between photon gas and dye solution is
achieved by reabsorption. To verify this experimentally we varied
the cut-off wavelength $\lambda_\text{c}=2\pi c/\omega_\textrm{c}$
by tuning the piezo voltage, as is shown in
Fig.~\ref{fig:figure7}. For smaller $\lambda_\text{c}$ the
absorption coefficient of the dye solution gets bigger and
consequently the emitted photons are reabsorbed faster. Thus a
sufficient thermal contact is expected to remain in this case. The
upper curves in Fig.~\ref{fig:figure7} confirm this (note the
logarithmic intensity scale). The thermal drop-off of the
transversal excitations is indeed visible. This does not hold
anymore if $\lambda_\text{c}$ is moved to a larger wavelength, see
the lower two spectra in Fig.~\ref{fig:figure7}. Here the
absorption coefficient is not big enough to achieve reabsorption
and consequently the spectra differ quite strongly from a
Boltzmann distribution. A similar observation can also be made by
varying the dye concentration. For low dye concentrations the
spectra are only partially thermalized.

\begin{figure}
\noindent \begin{centering}
\includegraphics[width=7cm]{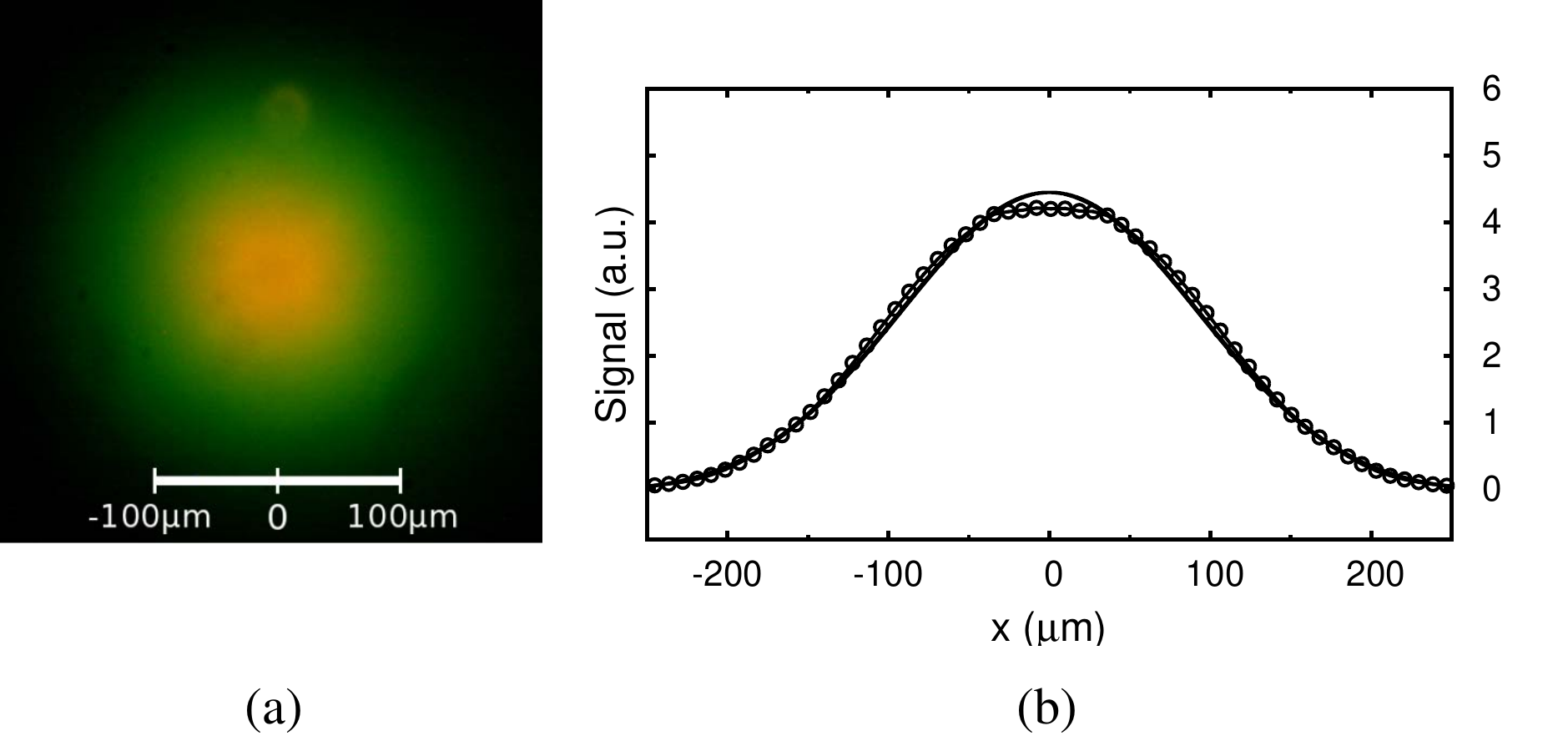}
\par\end{centering}

\caption{\label{fig:figure8}(a) Picture of the photon gas (real
image onto the sensor of a color CCD camera). Low-energy photons
(yellow) are emitted from the direct vicinity of the trap centre,
while the emission of higher-energetic photons (green) occurs
outside of the centre. (b) Photon distribution along an axis
intersecting the trap centre (circles), extracted from
Fig.~\ref{fig:figure8}(a). The theoretical distribution function
(line) is based on a thermal averaging over all resonator modes
and it corresponds to a Gaussian distribution. (Experimental
parameters as in Fig.~\ref{fig:figure6}) }

\end{figure}

Not only the spectral but also the spatial photon distribution
shows the expected characteristics of a thermal gas. In
Fig.~\ref{fig:figure8}a a picture of the photon gas taken by a
color CCD camera is shown. For that the photon gas was mapped with
a lens system as a magnified real image onto the CCD chip.
Transversally low excited photons (yellow) are seen in the direct
vicinity of the trap centre while the highly excited photons
(green) oscillate stronger in the transversal plane. From this
picture the spatial distribution of the photons, here a cut along
an axis through the centre, can be determined, see
Fig.~\ref{fig:figure8}b. Additionally the theoretical thermal
expectation value, which far below criticality is given by a Gauss
distribution, is plotted here (solid line). The good agreement
between the two curves again confirms the thermal character of the
photon gas. This is however not unexpected as the spectral
distribution of the photons has already revealed Boltzmann-like
statistical weights and spectral and spatial distributions can be
mapped onto each other in a thermalized system.

\begin{figure}
\noindent \begin{centering}
\includegraphics[width=7cm]{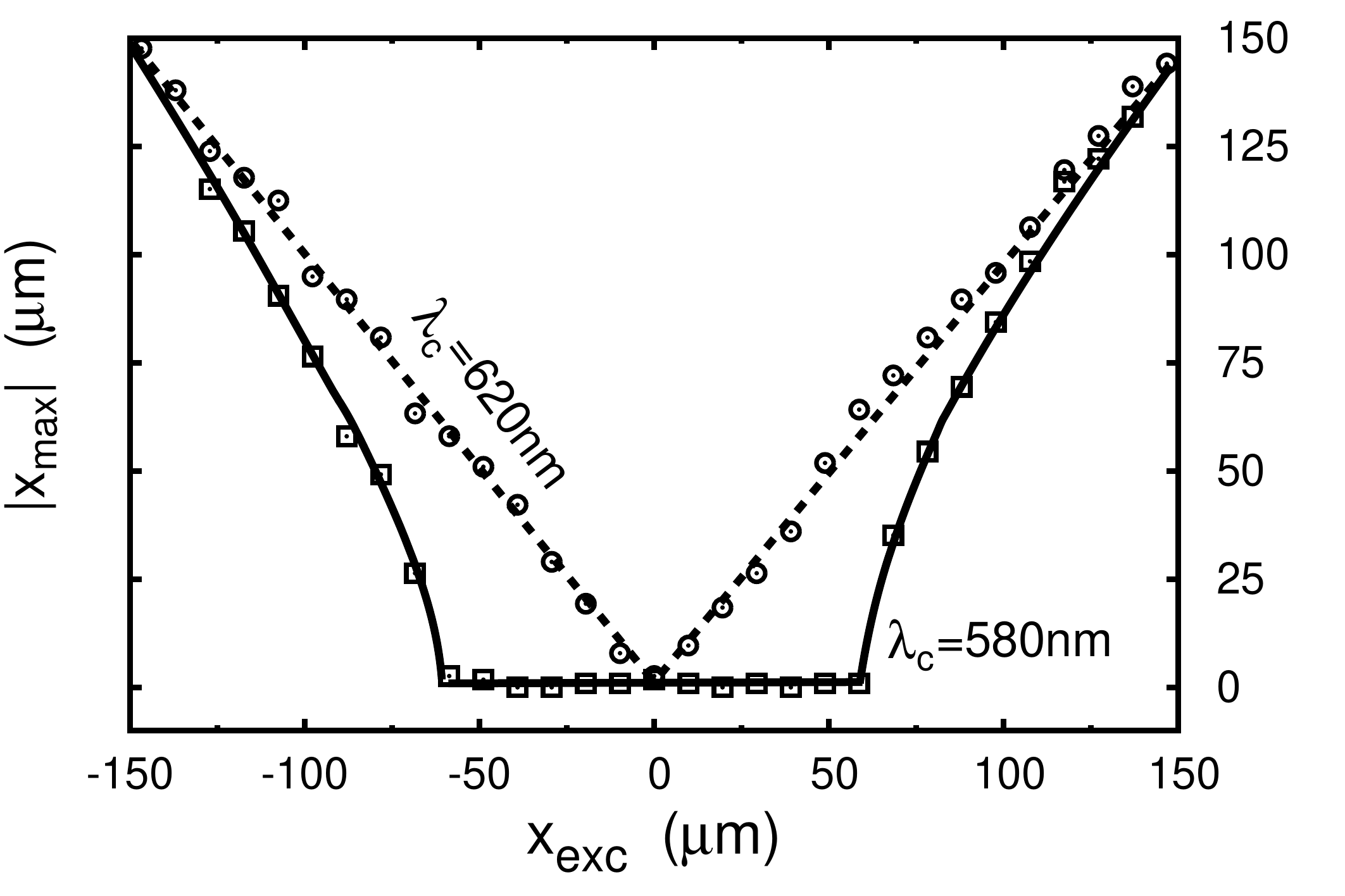}
\par\end{centering}

\caption{\label{fig:figure9}Measured distance of the intensity
maximum of the emitted fluorescence from the trap centre,
$|x_{\textrm{max}}|$, as a function of the position of the
excitation spot, $x_{\textrm{exc}}$. For a cut-off wavelength of
$\lambda_\text{c}=620\,\textrm{nm}$, corresponding to weak
reabsorption, the fluorescence exactly follows the pump spot
(circles connected by dashed line), for
$\lambda_\text{c}=580\,\textrm{nm}$, corresponding to strong
reabsorption, the photons concentrate to the trap centre,
$x_{\textrm{max}}=0$, provided the pump spot is displaced less
than $60\,\textrm{\textmu m}$ from the trap centre (squares
connected by a solid line). (PDI in acetone,
$\varrho=0.75\,\textrm{g/l}$, mirror curvatures
$R_{1}=R_{2}=1\,\textrm{m}$, $q=7$) }

\end{figure}

\subsection{Redistribution of photons}

To further characterize the thermalization process we investigated
the influence of the pump beam position. For that the pump beam is
displaced with respect to the optical axis. Due to the
thermalization process we nevertheless expect the photons to
accumulate in the trap centre where their potential energy is
minimized. The result of such a measurement can be found in
Fig.~\ref{fig:figure9}. In this figure the distance of the
fluorescence maximum from the optical axis as a function of the
pump position is shown. The pump spot is stepwise moved along a
line through the centre and at every position $x_{\textrm{exc}}$
the distance between the point of brightest fluorescence and the
optical axis $|x_{\textrm{max}}|$ is measured. The diameter (fwhm)
of the pump spot is roughly $25\,\textrm{\textmu m}$. These
measurements are performed for two different cut-off wavelengths
$\lambda_\text{c}$. At $\lambda_\text{c}\approx620\,\textrm{nm}$
reabsorption of the fluorescence light is nearly absent due to the
much smaller absorption coefficient in this spectral regime and
the decreasing mirror reflectivity. Consequently the photons leave
the cavity in the direct vicinity of the pump spot and the
opposite turning point, i.e. pump spot and absolute value of the
position of the intensity maximum are nearly identical. For
$\lambda_\text{c}\approx580\,\textrm{nm}$ reabsorption is strong
enough to trigger the thermalization process. Here the majority of
photons can accumulate in the potential minimum before they get
lost. However, this only holds if the position of the pump beam is
not too far away from the optical axis. This maximum distance is
determined to be roughly $\simeq60\,\textrm{\textmu m}$ for the
given experimental conditions. If the pump spot is moved beyond
that radius, the photon gas does not reach its equilibrium spatial
distribution anymore. The loss of spatial thermalization is
accompanied by a non-thermal spectral distribution. We interpret
this observed spatial redistribution effect as further evidence
for the presence of a thermalization process.

The reason for the incompleteness of the thermalization process
for stronger displaced pump spots is attributed to the finite
number of emission-reabsorption cycles a photon can undergo on
average before it is lost. For larger displacements this number
apparently is insufficient to reach the equilibrium spatial
distribution of the photon gas. We expect the most important loss
channel to be the coupling to unconfined modes. The mirrors used
in the experiments show a drop-off in reflectivity for angles
larger than $45\text{\textdegree}$. At every emission process
therefore a non-negligible probability exists that a photon is
scattered out of the cavity. Mirrors with a full three dimensional
bandgap would certainly improve on this
\cite{Noda:2000p1715,Blanco:2000p1730}. A second loss channel is
radiationless deactivation. The average number of
emission-reabsorption cycles is bounded by the quantum efficiency
smaller than unity. This average is given by
$\sum_{n=0}^{\infty}\Phi^{n}(1-\Phi)n=\Phi/(1-\Phi)$. For
rhodamine 6G ($\Phi_{\textrm{R6G}}=0.95$) it corresponds to $19$
fluorescence processes, for PDI ($\Phi_{\textrm{PDI}}=0.97$) it is
$32$. Both numbers seem to be rather big and we therefore do not
expect the finite quantum efficiency to be the dominant loss
mechanism. This is especially true because the quantum
efficiencies could even improve compared to the free space values
we have stated above. Such an effect is expected if the lifetime
of the excited molecules is shortened due to modified spontaneous
emission within the cavity.

\section{Bose-Einstein condensation}\label{sec:BEC}

\subsection{\label{sub:Subsection-title-1}Spectral and spatial photon distribution }

We now describe the behaviour of the photon gas for higher
intracavity powers. For these experiments the resonator is pumped
with powers of $P_{\textrm{pump}}\gtrsim100\,\textrm{mW}$. However
during the one-time pass of the pump beam through the cavity only
a small fraction of the light is actually absorbed in the dye film
\nobreakdash- less than $1\,\textrm{mW}$, the remainder is simply
transmitted. To prevent excessive population of dye triplet states
and heat development, the pump light is chopped into rectangular
pulses of $0.5\,\textrm{\textmu s}$ duration. The duty cycle is
typically 1:16000, i.e. every pulse is followed by a dark phase of
$8\,\textrm{ms}$ duration. For pulse durations longer than
$0.5\,\textrm{\textmu s}$ the intracavity power is observed to
decay. We attribute this to the population of triplet states. This
time scale is indeed also known for dye lasers as long as no
triplet quenchers are used \cite{Pappalardo:1970p1309}. Since the
pulses are at least two orders of magnitude longer than the
lifetime of the excited state dye molecules and roughly four
orders of magnitude longer than the lifetime of the photons, the
experimental conditions can be considered quasi-static.

\begin{figure}
\noindent \begin{centering}
\includegraphics[width=0.5\textwidth]{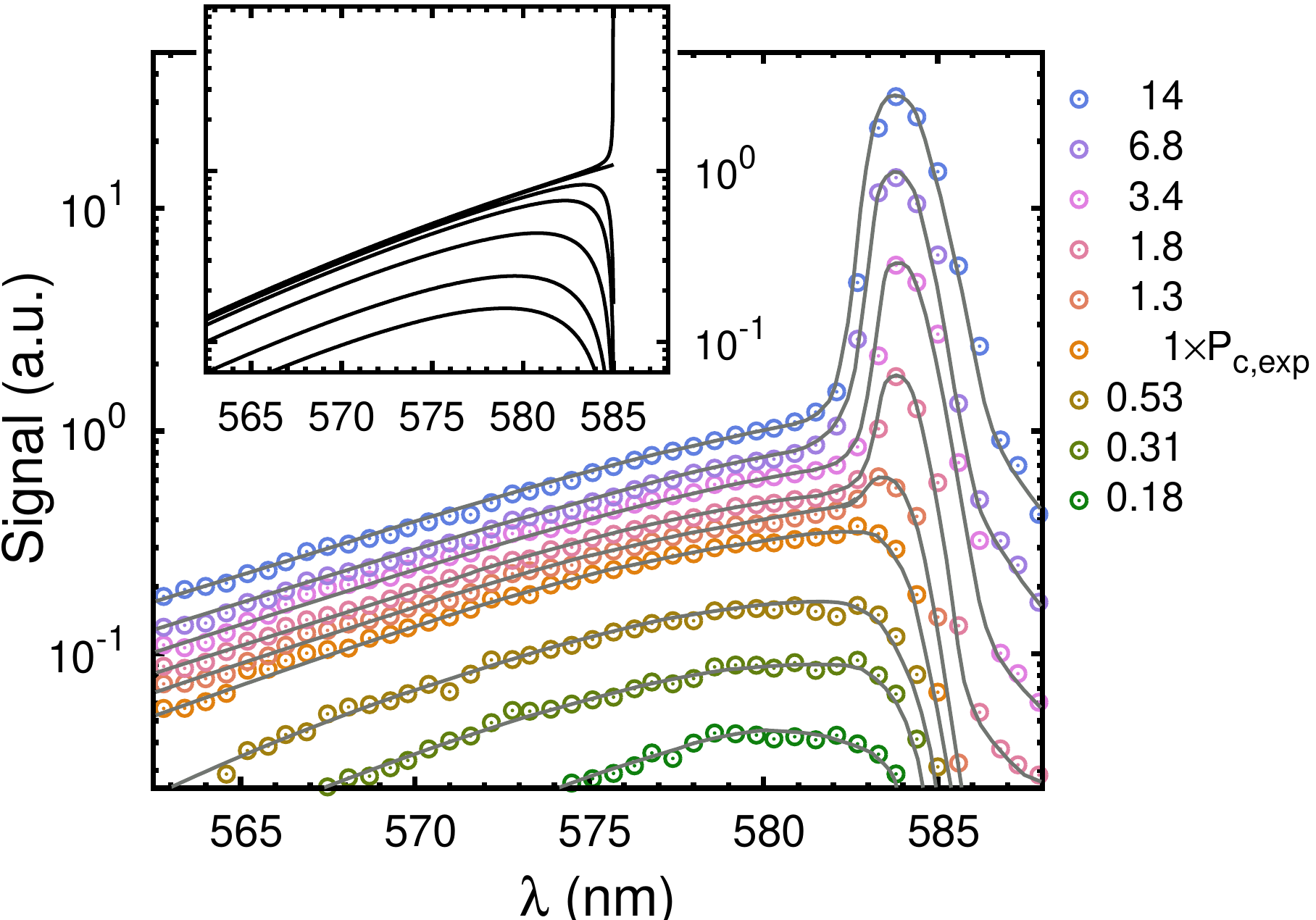}
\par\end{centering}

\caption{\label{fig:figure10}Spectral photon distribution for
increasing intracavity power. The optical powers are normalized to
the experimentally determined critical power
$P_{\textrm{c,exp}}=(1.55\pm0.60)\,\textrm{W}$, which corresponds
to a critical photon number of $N_{c}=(6.3\pm2.4)\times10^{4}$
(Rhodamine 6G in methanol,
$\varrho=1.5\times10^{-3}\,\textrm{Mol/l}$, mirror curvatures
$R_{1}=R_{2}=1\,\textrm{m}$, $q=7$, pulse duration
$0.5\,\textrm{\textmu s}$). The inset features theoretical spectra
based on a Bose-Einstein distribution of the transversal
excitations. }

\end{figure}

\begin{figure*}
\noindent \begin{centering}
\includegraphics[width=13cm]{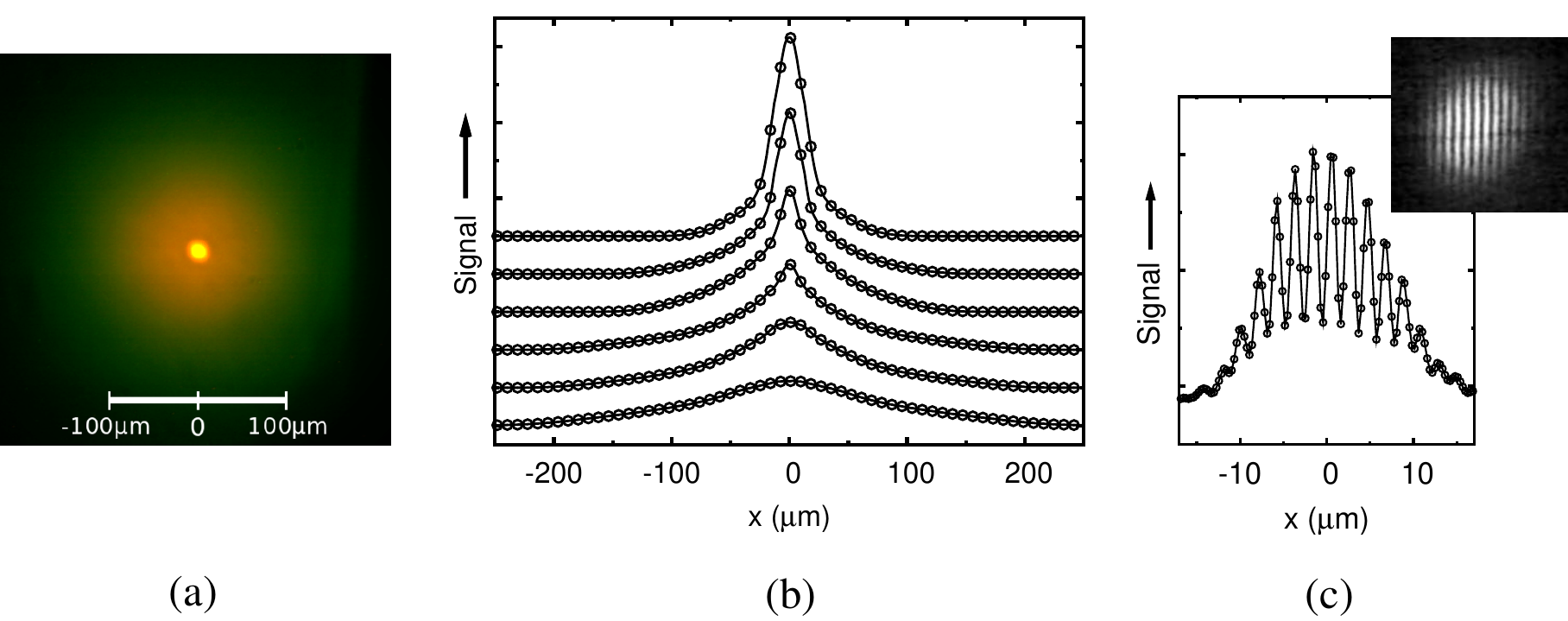}
\par\end{centering}

\caption{\label{fig:figure11}Picture of the photon gas (real image
onto the sensor of a color CCD camera). Low-energy photons
(yellow) are emitted from the direct vicinity of the trap centre,
while the emission of higher-energetic photons (green) occurs
outside of the centre. \label{fig:Raeumliche_Verteilung_2}(b)
Photon distributions along an axis intersecting the trap centre
for increasing order parameter $N_0/N\approx
\{0\%,0\%,1\%,3\%,10\%,25\%\}$ (circles). (c) Interference pattern
obtained by superimposing two partial beams of the condensate
using a Michelson-interferometer. Here the path difference of the
two interferometer arms is $15\,\textrm{mm}$. (experimental
parameters as in Fig.~\ref{fig:figure10}) }

\end{figure*}

The experimentally observed spectral distributions of the cavity
photons for increasing intracavity power is shown in
Fig.~\ref{fig:figure10}. In the inset the theoretical spectra
based on Bose-Einstein distributed transversal excitations at
$300\,\textrm{K}$ are depicted. The circulating light power inside
the cavity is determined by measuring the transmitted power
through one of the mirrors and using the transmission coefficient
as the proportionality factor. As observed before, the spectral
distribution is Boltzmann-like for low photon numbers. However for
higher photon numbers the maximum of the distribution slightly
shifts to larger wavelengths accompanied by a spectral narrowing.
Beyond a certain critical value, a spectrally sharp peak near the
cut-off appears. Due to the insufficient resolution of the
spectrometer of $\approx2\,\textrm{nm}$, it is not possible to
decide from the spectra which mode is massively populated.
Clear-cut evidence for the massive population of the ground mode
is however provided by the spatial distribution shown in
Fig.~\ref{fig:figure11}. The central bright spot visible has a
diameter (fwhm) of $(14\pm2)\,\textrm{\textmu m}$. This
corresponds well to the expected diameter of the transversal
ground state
$d_{\textrm{TEM}_{\textrm{q00}}}=2\sqrt{\hbar\ln2/m_{\textrm{ph}}\Omega}\simeq12.2\,\textrm{\textmu
m}$ for the given experimental parameters. The experimentally
determined critical light power inside the cavity is
$P_{\textrm{c,exp}}\approx(1.55\pm0.60)\,\textrm{W}$. If this
value is normalized by the power per photon,
$P_{\textrm{ph}}\simeq\hbar\omega_{0}/\tau_{\textrm{rt}}$, where
$\tau_\textrm{rt}$ is the round trip time, one obtains a critical
photon number of $N_{\textrm{c,exp}}\simeq
P_{\textrm{c,exp}}\tau_{\textrm{rt}}/\hbar\omega_{0}\approx(6.3\pm2.4)\times10^{4}$.
Within the quoted uncertainty, the value is in agreement with the
theoretically expected number of
$N_{\textrm{c}}=(\pi^{2}/3)(k_\textrm{B}T/\hbar\Omega)^{2}\simeq77000$
for the given parameter and thus confirms that the observed
condensation phenomenon occurs at a particle number which is
characteristic for a BEC. We have repeated the measurement for a
different dye (perylene-diimide (PDI),
$0.75\,\textrm{g}/\textrm{l}$ in acetone). Since the only property
of the dye that effects the condensation process should be its
temperature, we do not expect the critical photon number to differ
from the rhodamine case. This is indeed confirmed experimentally;
within the given uncertainties no deviation of
$P_{\textrm{c,exp}}$ or $N_{\textrm{c,exp}}$ is observed. There is
however a difference regarding the pump power that is needed to
achieve the critical photon number inside the resonator. The
reason for this is the lower absorption coefficient of PDI at the
pump wavelength of $532\,\textrm{nm}$. This is roughly a factor of
$3$ smaller than for rhodamine. To reach the same amount of
absorbed pump power one therefore has to pump harder.

The relatively large uncertainties regarding $P_{\textrm{c,exp}}$
and $N_{\textrm{c,exp}}$ stem from a combination of mainly
systematic error sources, especially uncertainties in the
calibration of the power meter and in the transmission coefficient
of the mirrors, $t=(2.5\pm0.4)\times10^{-5}$. Moreover, since the
cw-power coupled out at the phase transition is only of order
$\approx1\,\textrm{nW}$, care has to be taken that the power
measurement is not falsified by scattered pump light or background
fluorescence from the edges of the mirrors. Here a spatial
filterering has proved to be helpful. However, one has to ensure
that especially the higher transversal modes are not blocked by
this filtering. At last it is also necessary that the spatial pump
distribution, here a Gaussian ground mode with beam waist
$2w_{0}\approx100\,\textrm{\textmu m}$ centered around the optical
axis, differs not too strongly from the equilibrium distribution
of molecular excitations. As it already has been shown in the
previous section, the photon gas does not reach its equilibrium
state if this spatial mismatch gets too big. In this case a
position dependence of the chemical potential remains that is not
completely removed by the thermalization process before photons
leave the cavity. Such a non-equilibrium situation will also
influence the critical photon number.

To obtain the spatial photon distribution, pictures of the photon
gas for various pump levels have been recorded and analyzed
quantitatively. The intensity profiles (cut along an axis through
the trap center) shown in Fig.~\ref{fig:figure11}b have been
normalized to equal area and are shifted in the y-direction for
the sake of visibility. These profiles show an increasing spatial
concentration of the photons in the trap centre prior to the phase
transition \nobreakdash- a feature, which is in good agreement
with the theoretically calculated intensity profiles. However,
there is an interesting deviation of the experimental profiles
from the theoretical ones; the experiments reveal an increase of
the ground mode diameter with the ground state occupancy. This is
not expected for an ideal gas of photons and thus indicates the
presence of a repelling photon-photon interaction induced by the
medium. Such an interaction could e.g.\ arise from a
Kerr-nonlinearity of the dye molecules. However, in our case it
seems to stem mainly from a thermo-optical effect, further
investigations are presented in section 5.5. Another indication of
the selfinteraction is the non-perfect saturation of the thermal
modes, which can be attributed to a change in the effective
trapping potential for higher occupation numbers
\cite{Tammuz:2011arxiv}.

Preliminary experiments regarding the coherence of the condensate
have been performed. The light beam coupled out of the resonator
is sent through a Michelson interferometer with a pathway
difference of roughly $15\,\textrm{mm}$, see
Fig.~\ref{fig:figure11}c. For this measurement the interferometer
is slightly misaligned such that the two interfering beams do not
propagate collinear. The resulting interference pattern
demonstrates that the first order coherence of the condensate at
least extends into the centimeter regime. However, a more thorough
characterization of the coherence length is subject to ongoing
experiments and will be presented in a subsequent publication.

\subsection{Scaling of the critical power with the resonator geometry}

\begin{figure}
\noindent \begin{centering}
\includegraphics[width=8.5cm]{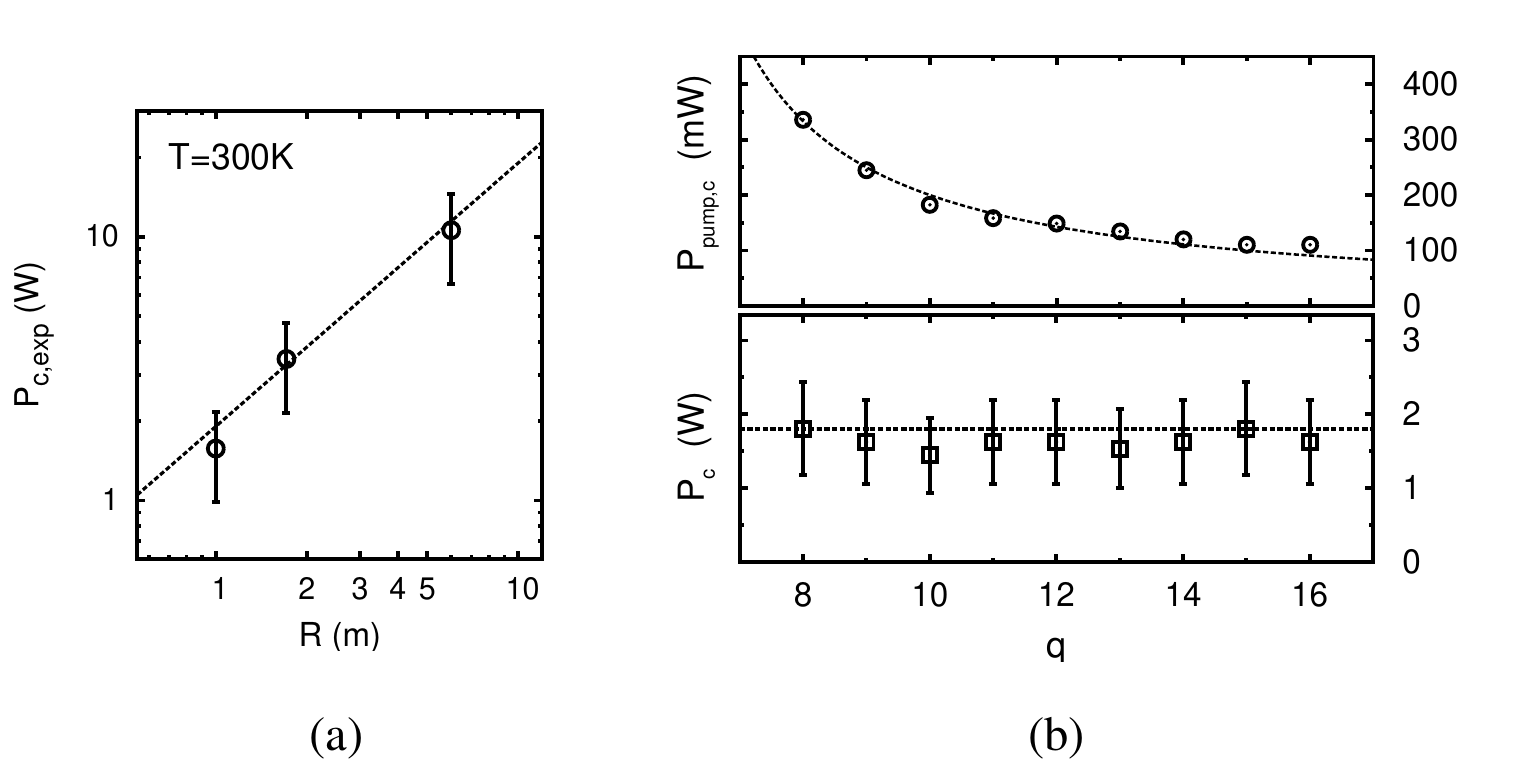}
\par\end{centering}

\caption{\label{fig:figure12}(a) Critical circulating intracavity
power as a function of the effective mirror curvature $R$.
(Rhodamine 6G in methanol,
$\varrho=1.5\times10^{-3}\,\textrm{Mol/l}$, $q=7$, pulse duration
$0.5\,\textrm{\textmu s}$) (b) Pump power at the phase transition
(top) and critical intracavity power (bottom), respectively, as a
function of the cavity order $q$. (Rhodamine 6G in methanol,
$\varrho=1.5\times10^{-3}\,\textrm{Mol/l}$, mirror curvatures
$R_{1}=R_{2}=1\,\textrm{m}$, pulse duration $0.5\,\textrm{\textmu
s}$) }

\end{figure}

The critical circulating power as a function of the geometry
parameters of the resonator can be casted in the form

\begin{equation}
P_{\textrm{c}}\simeq\frac{\pi^{2}}{12}\frac{n_{0}\omega_{0}}{\hbar
c}(k_{\textrm{B}}T)^{2}\, R.\label{eq:kritische
Leistung}\end{equation} One expects a linear increase of
$P_\textrm{c}$ with increasing radius of curvature $R$ and no
dependence on the longitudinal mode number $q$ \nobreakdash- at
least as long as the two-dimensionality of the photon gas holds.
To test the dependence on $R$, mirrors with two different
curvatures are available, $R_{1}=1\,\textrm{m}$ and
$R_{2}=6\,\textrm{m}$, allowing three different mirror
combinations. The combination $R_{\textrm{left}}=1\,\textrm{m}$
and $R_{\textrm{right}}=6\,\textrm{m}$ is equivalent to a
resonator built from two identical mirrors with a radius given by
the harmonic mean
$R=\,2\,(R_{1}^{-1}+R_{2}^{-1})^{-1}\simeq1.71\,\textrm{m}$. In
Fig.~\ref{fig:figure12}a the experimentally determined values of
the critical power as a function of the effective radius $R$ are
shown together with the theoretical expectation given by
eq.~\eqref{eq:kritische Leistung}. The data points coincide well
with the theoretic prediction \nobreakdash- both the linear
scaling and the absolute values. Seen in another way, the
procedure of increasing both the radius of curvature $R$ and the
photon number $N$ but preserving the ratio $R/N$ is observed to
retain the critical temperature of $T_{c}=300\,\textrm{K}$. Thus,
if this procedure was continued indefinitely, one could reach the
thermodynamic limit for condensation at a finite temperature. In a
further measurement we have investigated the dependence of the
critical power $P_{\textrm{c}}$ on the longitudinal mode number
$q$. For that, $q$ is increased stepwise without varying the
cut-off wavelength $\lambda_\text{c}=585\,\textrm{\textrm{nm}}$,
see Fig.~\ref{fig:figure12}b. As expected, $q$ has no significant
influence on $P_{c}$ within the given experimental uncertainties.
Both measurements again confirm that the observed condensation
occurs at the BEC criticality condition.

Interestingly, transitions to lower longitudinal modes were not
observed to be significant, which we attribute to the vanishing
spatial overlap of the modes. The observed spectra indicate that
the photon gas remains to good approximation two-dimensional up to
a cavity length of at least $16$ halfwaves. The pump power that is
necessary to achieve the critical photon number shows a strong
dependence on the mirror separation, see upper graph in
Fig.~\ref{fig:figure12}b. If the longitudinal mode number is
doubled from $q=8$ to $q=16$ by increasing the mirror separation,
only one third of the initial pump power is necessary to reach
criticality. Thus the system shows a behaviour contrary to
threshold behaviour reported for dye
\cite{DeMartini:1988p632,DEMARTINI:1992p683} or semiconductor
microlasers \cite{Yokoyama:1992p2123}, where an increase of the
threshold pump power with the cavity length has been observed. The
reason for the here observed decrease of the pump power is an
increase of the absorbance for thicker dye films. Since the
observed critical circulating power $P_{c}$ is independent of the
cavity length, the absorbed pump power at criticality should also
be roughly constant. Thus in the weak absorption limit, as in our
case, one expects to see a reciprocal dependence of the pump power
$P_{\textrm{pump}}$ on the dye film thickness
$D_{\textrm{dye}}(q)$, as this conserves the amount of absorbed
pump power. Thus we expect
$P_{\textrm{pump}}\propto(q-q_{0})^{-1}$, with the penetration
depth into the mirrors $q_{0}$. The corresponding fit to the
experimental data is shown in the upper graph of
Fig.~\ref{fig:figure12}b. For the parameter $q_{0}$ one obtains a
value of $q_{0}=4.77\pm0.25$, in good agreement with the value of
$q_{0}=4.68\pm0.17$ derived in section 4.1 by directly measuring
the absorbance. This confirms that the necessary absorbed pump
power to trigger criticality remains constant for various mirror
separations. The given values of the critical pump power can also
be converted to pump intensities,
$I_{\textrm{pump,c}}=P_{\textrm{pump,c}}/\pi w_{0}^{2}$, where
$w_{0}\simeq50\,\textrm{\textmu m}$ is the pump beam radius. One
obtains values roughly in the regime
$(0.6-2.2)\,\textrm{kW/cm}^{2}$, where the given values of
$P_{\textrm{pump},c}$ in Fig.~\ref{fig:figure12}b have to be
corrected by the mirror transmission coefficient, which is of
order $50\%$. These values are roughly two orders of magnitude
smaller than in macroscopic dye lasers, where typically pump
intensities of order $\approx100\,\textrm{kW/cm}^{2}$ are required
\cite{Ippen:QuantumElectronics1971,Peterson:ApplPhysLett1970}.

\subsection{Condensation by spatial relaxation}

As it was already demonstrated in section 4.2, the thermalization
process is accompanied by a spatial redistribution of the photons
towards the trap centre. This redistribution can also be observed
for higher photon numbers. It can even provide a sufficiently high
density at the trap centre to trigger the condensation. To
demonstrate this, we have performed the following experiment: the
dye solution is pumped $50\,\textrm{\textmu m}$ away from the
optical axis with a pump beam of roughly $35\,\textrm{\textmu m}$
(fwhm) diameter. Both its position and its power are kept fixed in
the course of the experiment. By fine tuning the cavity cut-off
$\lambda_\text{c}$, we can vary the amount of reabsorption.

\begin{figure}
\noindent \begin{centering}
\includegraphics[width=6.25cm]{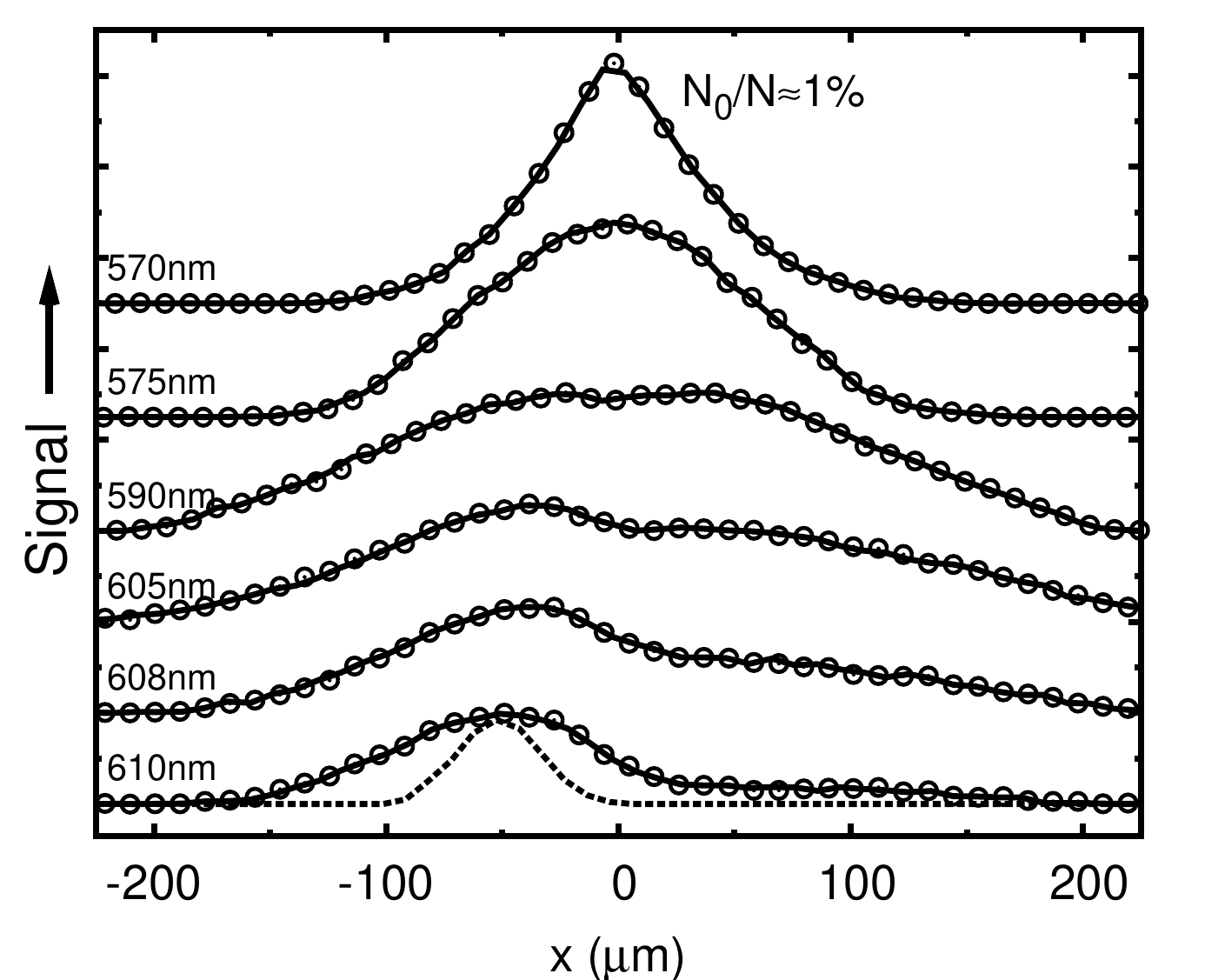}
\par\end{centering}

\caption{\label{fig:figure13}Intensity distribution of the photon
gas along an axis intersecting the trap centre for varied cut-off
wavelengths $\lambda_\text{c}$ (left edge). The pump beam (dashed
line) is located outside the trap centre and its position as well
as its power are kept fixed during the measurement. The top curve
shows a photon gas precisely at the phase transition with a ground
state population $N_{0}/N\lesssim1\%$. (Rhodamine 6G in methanol,
$\varrho=1.5\times10^{-3}\,\textrm{Mol/l}$, mirror curvatures
$R_{1}=R_{2}=1\,\textrm{m}$, $q=7$, pulse duration
$0.5\,\textrm{\textmu s}$) }

\end{figure}

The lowest profile in Fig.~\ref{fig:figure13} shows the intensity
of the photon gas along a line through both trap centre and pump
position for a cavity cut-off of
$\lambda_\text{c}=610\,\textrm{nm}$. Additionally the profile of
the pump spot alone is displayed, where thereto one of the cavity
mirrors was removed. For this value of $\lambda_\text{c}$ the
degree of reabsorbtion is low and the thermalization process is
suppressed. The observed spatial distribution of the photon gas
roughly follows the pump beam intensity distribution. If
$\lambda_\text{c}$ is decreased, the amount of reabsorption is
increased and the thermal contact between photon gas and dye
solution is gradually established. The intensity profiles get
distributed more symmetrically around the trap centre. For a
cavity cut-off of $\lambda_\text{c}=570\,\textrm{nm}$ the photon
density in the centre even gets sufficiently high to trigger
criticality. This is both indicated by a small bright spot
occurring at the position of $\textrm{TEM}_{00}$-mode and the
corresponding cusp in the intensity profile (upper curve in
Fig.~\ref{fig:figure13}). However, the condensate fraction that is
achieved in this way is rather low, $N_{0}/N\lesssim
1\textrm{\%}$.

An important conclusion from this measurement is that the
condensation is triggered not directly by the pumping. The
intensity of the pump beam at the position of the
$\textrm{TEM}_{00}$-mode is almost negligible. We attribute this
to the multiple emission-reabsorption cycles, the induced
thermalization process provides a sufficient photon density for
the condensation. Such an effect is unknown in lasers but has
already been observed in the context of polariton
condensation\cite{Balili:2007p1342}. As in section 4.2, a complete
spatial relaxation of the photon gas is achieved if the pump spot
is not moved beyond a distance of $\simeq50\,\textrm{\textmu m}$
from the trap centre. Again this can be attributed to the finite
number of emission-reabsorption cycles a photon can undergo before
it is lost.

\subsection{Reabsorption cycles}

Perhaps the most decisive quantity regarding the degree of
thermalization is the average number of reabsorptions of a photon
before it is lost from the resonator. The observed spatial
redistribution of the photons towards the trap centre already
tells us qualitatively that multiple reabsorptions take place in
the experiment. A more quantitative measure is experimentally
obtainable by a time-resolved measurement, for example of the
overall decay time of the microresonator fluorescence. Such a
measurement is under current preparation. Nevertheless, already an
overall input-output analysis of the data presented in
Fig.~\ref{fig:figure12}b confirms that the experiments are
performed in a regime of multiple reabsorptions, and gives an
upper bound for the number of reabsorptions per photon.

From the general conservation of excitations we conclude that the
average numbers of excited molecules $N_{\textrm{exc}}$ and the
number of photons $N_{\textrm{ph}}$ are linked by
\begin{equation}
N_{\textrm{exc}}\,\tau_{\textrm{exc}}^{-1}\,\beta=N_{\textrm{ph}}\,\tau_{\textrm{ph}}^{-1},\label{eq:nonetflux}\end{equation}
where $\tau_{\textrm{ph}}$ is the average time between emission
and reabsorption of a photon, and $\tau_{\textrm{exc}}$ the
lifetime of the molecular electronic excitation in the resonator.
Further, $\beta$ is the probability that an excited molecule emits
into a confined cavity mode rather than decaying radiationless or
into an unconfined mode. With this equation we implicitly make the
approximation that losses predominantly occur for conversions from
molecular excitations to cavity photons but not for the inverse
direction, which seems to be justified in our system.
$N_{\textrm{exc}}$ is also linked to the absorbed pumping power
$P_{\textrm{p,abs}}$ by\begin{eqnarray}
N_{\textrm{exc}} & = & \frac{P_{\textrm{p,abs}}}{h\nu_{\textrm{p}}}\,\tau_{\textrm{exc}}\,(1+\beta+\beta^{2}+\ldots)\nonumber \\
 & = & \frac{P_{\textrm{p,abs}}}{h\nu_{\textrm{p}}}\,\tau_{\textrm{exc}}\,\frac{1}{1-\beta},\label{eq:N_exc}\end{eqnarray}
where $h\nu_{\textrm{p}}$ is the energy of a pump photon and thus
$P_{\textrm{p,abs}}/h\nu_{\textrm{p}}$ is the absorbed pump photon
rate. The fraction of excitations that were directly created by
the pump beam corresponds to the term $'1'$ of the sum, the
fraction of excitations that were reabsorbed once corresponds to
the term $'\beta'$, and so on. We here assume that the storage
time in the resonator is dominated by the lifetime of the
molecular excitation, $\tau_{\textrm{exc}}\gg\tau_{\textrm{ph}}$.
By the combination of eq.~\eqref{eq:nonetflux} und
eq.~\eqref{eq:N_exc} the probability $\beta$ can be
derived\begin{equation}
\beta=\frac{1}{1+\frac{P_{\textrm{p,abs}}}{h\nu_{\textrm{p}}}\frac{\tau_{\textrm{ph}}}{N_{\textrm{ph}}}},\end{equation}
and with that, the average number of reabsorptions per photon
$\bar{n}_{\textrm{re}}=\sum_{n=0}^{\infty}n\beta^{n}(1-\beta)$ is
written as:\begin{equation}
\bar{n}_{\textrm{re}}=\frac{h\nu_{\textrm{p}}}{P_{\textrm{p,abs}}}\frac{N_{\textrm{ph}}}{\tau_{\textrm{ph}}},\label{eq:n_re}\end{equation}
The lifetime $\tau_{\textrm{ph}}$ is experimentally controlled by
the dye concentration and is also strongly dependent on the cavity
cut-off $\lambda_\text{c}$. For the experimental parameters of
Fig.~\ref{fig:figure12}a we expect a value of
$\tau_{\textrm{ph}}=(21\pm6)\,\textrm{ps}$. The absorbed pump
power is estimated to be
$P_{\textrm{p,abs}}=(0.65\pm0.10)\,\textrm{mW}$. With that we
obtain $\bar{n}_{\textrm{re}}=3.8\pm2.5$. The error of this
estimate is rather big, nevertheless $\bar{n}_{\textrm{re}}$ is
clearly found to be of magnitude 1. The quantitative analysis of
the particle flux thus confirms that the experiment is carried out
in the regime of multiple scattering. However, it delivers an
upper bound of $\bar{n}_{\textrm{re}}\leq6$. At first sight this
number seems to be relatively small compared to the numerous
collisions taking place in an atomic gas. Regarding the degree of
thermalization this might seem unfavourable. But one has to
consider that, in contrast to atomic collisions, nearly all
correlations between absorbed and emitted photon states are
annihilated already by a single absorption-emission process
(Kasha's rule\cite{Kasha:1950,Lakowicz:1999}) \nobreakdash- with
the only exception of a necessary spatial overlap between both
photon states. Thus, the contact with a heat bath, as it takes
place in our photon gas experiment, is a much stronger
thermalization process compared to two-body collisions in an
atomic gas. Experimentally, we have strong evidence for the
thermalization to operate both properly in the spatial and the
spectral regime, see chapter \ref{sec:thermalization}.

\subsection{Selfinteraction of the light condensate}

\begin{figure}
\noindent \begin{centering}
\includegraphics[width=7.5cm]{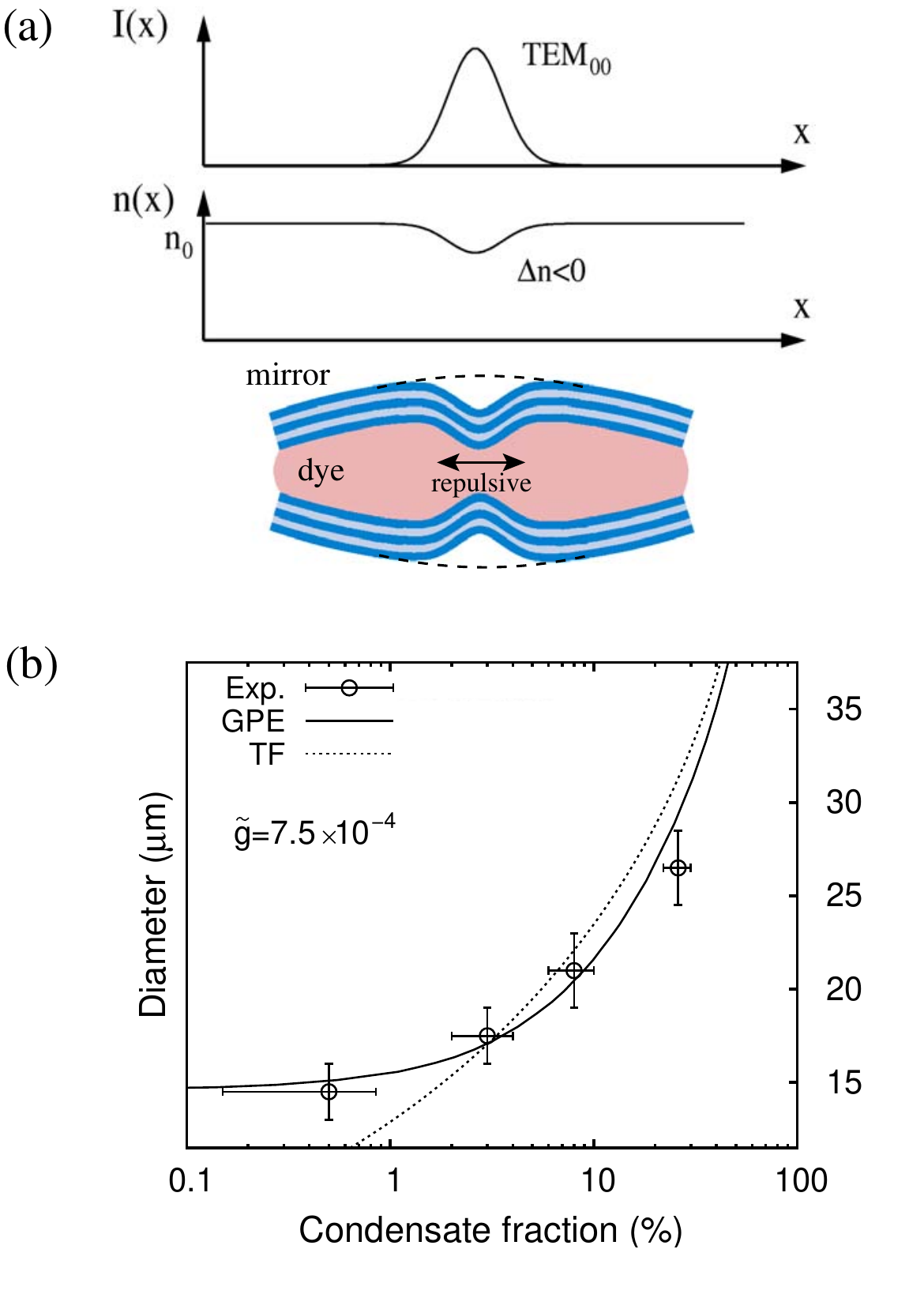}
\par\end{centering}

\caption{\label{fig:figure14}(a) Schematic illustration of the
optical self-interaction. The massively occupied ground state is
expected to locally heat up the dye solution, reducing the index
of refraction and, therefore, shortens the optical distance
between the mirrors. This is effectively equivalent to a local
deformation of the mirrors (which is shown exaggerated in the
figure), resulting in a repulsive interaction. (b) Diameter (fwhm)
of the fundamental transversal mode ($\textrm{TEM}_{00}$) as a
function of the condensate fraction (circles). The solid line
indicates the theoretically predicted mode diameter for an
interaction parameter of $\tilde{g}=7.5\times10^{-4}$ expected
from the Gross-Pitaevskii equation, the dashed line gives the
diameter expected from a Thomas-Fermi approximation. (experimental
parameters as in Fig.~\ref{fig:figure11}) }

\end{figure}

The observed spatial intensity distributions
(Fig.~\ref{fig:figure11}b) reveal an increase of the condensate
diameter with increasing occupation level. The diameter (fwhm) of
the transversal ground state as function of
$N_{0}/N_{\textrm{ph}}$ is given explicitly in
Fig.~\ref{fig:figure14}b. Slightly above criticality the measured
diameter of $(14\pm2)\,\textrm{\textmu m}$ is in good agreement
with the expected ground state diameter of\begin{equation}
d_{0}=2\sqrt{\hbar\ln2/m_{\textrm{ph}}\Omega},\label{eq:d0}\end{equation}
for non-interacting photons, yielding
$d_{0}\simeq14.7\,\textrm{\textmu m}$ for the parameters of
Fig.~\ref{fig:figure11}. For higher ground state occupation
$N_{0}/N_{\textrm{ph}}$ this does not hold anymore. This
observation suggests the presence of a repelling photon-photon
interaction mediated by the dye solution, which can formally be
described by a non-zero nonlinear index of refraction,
$n_{2}\neq0$, see eq.~\eqref{eq:E_ph}. In our case the dominant
contribution to $n_{2}$ seems to be given by a thermo-optical
effect. The massively occupied ground state should lead to a small
heat gradient in the dye solution due to nonradiative decays. This
heat gradient then causes a gradient of the refractive index,
which decreases the \textit{optical} distance between the mirrors
on the optical axis. Effectively, this can be interpreted as a
local distortion of the mirrors, see Fig.~\ref{fig:figure14}a,
which is indeed expected to lead to a broadening of the ground
state mode \nobreakdash- in qualitative agreement with the
experimental observation.

The index gradient necessary to notably broaden the condensate
peak can be estimated by comparing he transversal ground state
energy, $\hbar\Omega$, with the interaction term
$E_{\textrm{int}}$ of eq.~\eqref{eq:E_ph}
\begin{equation}
E_{\textrm{int}}=-\frac{m_{\textrm{ph}}c^{2}}{n_{0}^{3}}\Delta
n,\label{eq:E_int}\end{equation} with $\Delta n=n_{2}I(\vec{r})$.
They become equal for a gradient of $\Delta n\simeq-\hbar\Omega
n_{0}^{3}/m_{\textrm{ph}}c^{2}$ which in our experiment is roughly
$\Delta n\simeq-0.9\times10^{-4}$. Such a change in the refractive
index is caused by a temperature gradient of $\Delta T=(\partial
n/\partial T)^{-1}\Delta n\simeq0.17\,\textrm{K}$, where for
methanol we used $\partial n/\partial T\simeq - 4.86 \times
10^{-4}\,\textrm{K}^{-1}$  \cite{Lusty:1987p1307}, and thus we
expect thermal effects to play an important role. In principle,
higher orders of polarizability as e.g.~the optical Kerr-effect,
contribute to the nonlinear index of refraction. This is however
expected to be a much weaker process: for an index change of
magnitude $\Delta n=10^{-4}$, an intensity of $I\simeq\Delta
n/n_{2}^{\textrm{(Kerr)}}\simeq1\,\textrm{GW}/\textrm{cm}^{2}$ is
necessary, which is $4$ to $5$ orders of magnitude higher than the
intensity building up in the microresonator. Here a nonlinear
index of refraction of magnitude
$n_{2}^{\textrm{(Kerr)}}\simeq-1\times10^{-13}\,\textrm{cm}^{2}/\textrm{W}$
was assumed \cite{Nag:2009p2089}.

The interaction term, eq.~\eqref{eq:E_int}, can be cast into a
more familiar form by introducing the condensate wavefunction
$\psi_{0}(\vec{r})$ via
$I(\vec{r})=(m_{ph}c^{2}\,/n_{0}^{2}\tau_{\textrm{rt}})\,
N_{0}\left|\psi_{0}(\vec{r})\right|^{2}$, where
$\tau_{\textrm{rt}}=hqn_{0}^{2}/m_{\textrm{ph}}c^{2}$ is the
resonator round trip time. Then eq.~\eqref{eq:E_int} reads as
\begin{equation}
E_{\textrm{int}}=\frac{\hbar^{2}}{m_{\textrm{ph}}}\tilde{g}N_{0}|\psi_{0}(\vec{r})|^{2},\label{eq:E_int_2}\end{equation}
with the dimensionless interaction parameter $\tilde{g}$ given
by
\begin{equation}
\tilde{g}=-\frac{m_{\textrm{ph}}^{3}c^{4}n_{2}}{\hbar^{2}n_{0}^{5}\tau_{\textrm{rt}}}.\end{equation}
With the interaction term in the form of eq.~\eqref{eq:E_int_2}
the similarity of the photon energy given by eq.~\eqref{eq:E_ph}
to the Gross-Pitaevskii (GPE) equation becomes even more apparent.
To determine $\tilde{g}$ in our experiment we measure the
occupation level $N_{0}/N_{\textrm{ph}}$ at which the condensate
diameter is doubled compared to the non-interacting case,
eq.~\eqref{eq:d0}. This is achieved roughly at
$N_{0}/N\simeq0.25$, which corresponds to an occupation number of
$N_{0}\simeq40000$. From a numeric solution of the 2d GPE we
obtain a doubling of the diameter for an interaction strength of
$\tilde{g}N_{0}\simeq30$, giving the dimensionless interaction
parameter in our experiment,
$\tilde{g}\simeq30/40000=7(3)\times10^{-4}$.
Fig.~\ref{fig:figure14}b shows the experimentally determined data
points together with a curve based on the numerical solution to
the 2d GPE (solid line) where we used $\tilde{g}=7.5\times10^{-4}$
as an input parameter. Apparently the experimental data can be
modelled satisfactorily by this. Additionally, in the figure we
have included the condensate diameter (fwhm) according to a
Thomas-Fermi approximation (dashed line), given by
\begin{equation}
d_{0}^{\textrm{TF}}=2\sqrt{\hbar/\sqrt{\pi}m_{\textrm{ph}}\Omega}\,(\tilde{g}N_{0})^{\frac{1}{4}},\end{equation}
which is supposed to yield approximatively correct results for
higher condensate fractions, where the transversal kinetic energy
of the photons can indeed be neglected.

Compared to the interaction parameters reported for
two-dimensional atomic Bose gases, which are in the regime
$\tilde{g}=10^{-2}-10^{-1}$ \cite{Hadzibabic:2009p2114}, the
interactions in the photon gas are at least one order of magnitude
smaller. Thus the photon gas is closer to an ideal Bose gas, and
there is no indication that the observed phase transition differs
from the BEC scenario. This would be expected for stronger
interacting gases, where the interactions give rise to a
Kosterlitz-Thouless (KT) type of phase
\cite{Hadzibabic:2006,Hadzibabic:2009p2114,Clade:PRL2009}. An
indicator for this would be the loss of long range spatial order.
In initial experiments using a shearing interferometer, see
Fig.~\ref{fig:figure11}, we have tested for the first-order
coherence of the condensate by bringing different spatial parts of
the condensate to overlap. The observed interference patterns
however did not show any signatures of a spatially varying phase.
This is in agreement with theoretical expectations, predicting
that long range order is lost only for higher values of the
dimensionless interaction parameter \cite{Hadzibabic:2009p2114}.

\section{Conclusions}

We have investigated thermodynamic properties of a two-dimensional
photon gas in a dye-filled  optical microcavity. In initial
experiments, we have investigated spectral and spatial properties
of the photon gas below criticality in detail, and obtained
evidence for a thermalized two-dimensional photon gas with
non-vanishing chemical potential. When the photon number is
in\-creased above a critical value, Bose-Einstein condensation of
the photon gas is observed. Experimental signatures  are
Bose-Einstein distributed transversal excitations including a
macroscopic population of the transversal ground mode, the
condensation setting in at the expected critical photon number and
a spatial relaxation process into the trap centre that can lead to
condensation even for a displaced pump spot.

The hitherto performed experiments were focussed on the
thermodynamics of the photon gas, a more detailed investigation of
the condensate properties is currently performed. Initial
interferometric measurements of the condensate peak have already
confirmed the first-order coherence of the condensate, with a
lower bound for the coherence length of the order of centimeters.
We also plan to investigate the second-order coherence
\cite{HBT:1956p1046,Purcell:1956p1449}, which allows to monitor
the unusually large particle fluctuations expected from a
grand-canonical ensemble. This would allow to further distinguish
the radiation emitted by the equilibrium photon BEC from
nonequilibrium laser radiation.

On a longer timescale, an experiment based on photonic crystal
mirrors with a three-dimensional bandgap
\cite{Noda:2000p1715,Blanco:2000p1730} should prove advantageous,
as losses due to spontaneous emission out of the cavity should be
reduced. This would allow to reduce the amount of pumping to
maintain the average photon number in the resonator. The full
bandgap mirrors together with a condensed matter system, e.g.~dye
molecules in a polymer matrix, could even prove useful for
technical applications, as it would allow the construction of
fluorescence collectors \cite{VanSark:2008p727}, that are more
efficient than currently used ones. A tantalizing perspective
includes the possible realization of coherent light sources in the
UV-regime. Other than in a laser, no inversion is required for the
realization of a photon BEC, and spontaneous emission is
recaptured
\cite{DeMartini:1988p632,Yokoyama:1992p2123,Yamamoto:1992p2125}.

On the fundamental research side, we expect the photonic gas
within a 'white-wall' box to be a versatile tool. For the ideal
Bose gas one does not expect superfluidity, in contrast to the
interacting case. Therefore it would be interesting to observe the
superfluid properties of the condensate as a function of the
dimensionless interaction parameter $\tilde{g}$. A possible route
for the tuning of $\tilde{g}$ would be either the tuning of the
photon mass by moving the experiment to the UV spectral region,
the reduction of the resonator length by reducing the penetration
of the electric field into the mirrors, and by increasing the
nonlinear coefficient $n_2$. We expect that the interaction
parameter of currently $\tilde{g}\approx7\times10^{-4}$ can be
increased by at least an order of magnitude, thus allowing to
investigate the superfluid properties over a wide range of
parameters. We are aware that the character of the interaction
needs to be investigated further. In a different line of
experiments, the usage of a second 'heating' dye molecule with low
quantum efficiency, whose absorption is spectrally shifted away
from the observed spectral region, should allow to thermooptically
change the refractive index by means of a heating laser. This
would allow the modelling of e.g.~periodic lattices, and also more
complex potentials for the photon gas, by simply imaging the
corresponding pattern onto the 'heating' dye molecules.

\begin{acknowledgements}
%If you'd like to thank anyone, place your comments here.
Financial support from the Deutsche Forschungsgemeinschaft within
the focused research unit FOR557 and under contract WE1748/09 is
acknowledged.

\end{acknowledgements}

\section*{---------------------}

\bibliographystyle{spphys}
\bibliography{Review_PBEC}

\end{document}